\documentclass[superscriptaddress,column,showpacs,a4paper,
amssymb,amsmath,nobibnotes,aps,prd,longbibliography,
showkeys,
nofootinbib,notitlepage]{revtex4-1}
\usepackage{bigints}
\usepackage{graphicx}
\usepackage{float}
\usepackage{epsf}
\usepackage{bm}
\usepackage{amsmath}
\usepackage{amsfonts}
\usepackage{amssymb}
\usepackage{epstopdf}
\usepackage{natbib}
\usepackage{color}%
\providecommand{\U}[1]{\protect\rule{.1in}{.1in}}
\newcommand{\be}{\begin{equation}}
\newcommand{\ee}{\end{equation}}

\newcommand{\mincir}{\raise
-3.truept\hbox{\rlap{\hbox{$\sim$}}\raise4.truept\hbox{$<$}\ }}
\newcommand{\magcir}{\raise
-3.truept\hbox{\rlap{\hbox{$\sim$}}\raise4.truept\hbox{$>$}\ }}

\newtheorem{remark}{Remark}[section]
\usepackage{xcolor}
\usepackage{hyperref}
\hypersetup{colorlinks=true, linkcolor=blue,citecolor=magenta}

\begin{document}

\title{{A dynamical approach to General Relativity based on proper time
}}

\author{Jaume de Haro}
\email{jaime.haro@upc.edu}
\affiliation{Departament de Matem\`atiques, Universitat Polit\`ecnica de Catalunya, Diagonal 647, 08028 Barcelona, Spain}

\begin{abstract}
This work places the invariant 
$ds^2$ at the center of the gravitational interaction, interpreting it not as a purely geometric object but as the differential of proper time, endowed with direct physical meaning.
Starting from the extension of Fermat’s principle to massive particles—namely, the requirement that freely falling bodies follow trajectories that extremize proper time, {which for timelike motion corresponds to a local maximum}—and invoking the universality of Galilean free fall, we derive the form of $ds^2$ in a static gravitational field.
Lorentz invariance then provides the natural framework to extend this result to systems involving moving matter. The invariant derived through this procedure matches the weak-field limit of General Relativity formulated in the harmonic gauge.

Within this linearized regime, we show that the structure of the theory already contains the seeds of its non-linear completion: any dynamically consistent extension to strong gravitational fields necessarily involves the Ricci tensor. From this viewpoint, Einstein’s field equations appear not as a postulated geometric law, but as the unique covariant closure required to ensure energy–momentum conservation and the self-consistency of the gravitational interaction.

\end{abstract}

\vspace{0.5cm}


\maketitle

\section{Introduction}

General Relativity stands as one of the most remarkable intellectual achievements of the twentieth century. Its significance lies not only in its empirical success, but also in the fact that it emerged largely from the sustained effort of a single visionary mind—Albert Einstein—who, with relatively few yet decisive collaborations, profoundly reshaped our understanding of space, time, and gravitation.

Within Newtonian physics, gravity appeared as a deeply puzzling interaction: an instantaneous force acting across empty space \cite{Newton}. Einstein’s insight, as it is commonly summarized, went far beyond this picture. Gravitation ceased to be interpreted as a force transmitted at a distance and was instead understood as a manifestation of the structure of spacetime itself \cite{Einstein1955}. Matter influences that structure, and bodies—whether massive particles or light rays—follow trajectories dictated by it. Although this imagery is often conveyed through evocative metaphors, it should be emphasized that such language does not fully capture Einstein’s mature viewpoint. As Einstein himself stressed, there is no physical medium that literally bends or deforms \cite{Einstein1920}; rather, curvature constitutes an efficient and powerful way of encoding the dynamical relations between matter, motion, and inertia.

In most pedagogical treatments and standard references \cite{MTW,Synge,Dirac,HE,Weyl,Eddington}, General Relativity is introduced by first establishing its mathematical framework. One begins with the machinery of differential geometry: metrics, affine connections—most notably the Levi--Civita connection—and curvature tensors such as the Riemann and Ricci tensors. Only after this formal structure is laid out are the central physical principles articulated: the Principle of Relativity, the Equivalence Principle, and the Principle of General Covariance. The energy--momentum tensor is then introduced as the mathematical representation of matter. With all these ingredients in place, Einstein’s field equations emerge as the link between spacetime geometry and material content, within a pseudo-Riemannian (or Lorentzian) geometric setting.

This conceptual synthesis is often encapsulated by John Archibald Wheeler’s celebrated dictum \cite{TW}:
\begin{quote}
\textit{Matter tells spacetime how to curve, and spacetime tells matter how to move.}
\end{quote}

Despite its elegance, this formulation raises a fundamental conceptual question: what does it truly mean to say that spacetime is curved? A familiar pedagogical analogy is frequently invoked: a heavy object placed on a stretched elastic sheet produces a depression, and a smaller object rolling nearby follows a curved path. This illustration aims to offer an intuitive picture of how paths may curve away from straight lines without requiring a force acting remotely.

However, this analogy is strictly illustrative and, upon closer inspection, misleading. It implicitly relies on the notion of weight—namely, on the Newtonian gravitational force exerted by the Earth—and therefore presupposes precisely the framework it aims to transcend. As such, it obscures rather than clarifies the genuine physical content of the theory.

What, then, carries true physical meaning in General Relativity? The answer cannot lie in coordinates, nor in the particular form of the metric expressed in those coordinates, but rather in invariant quantities. Among these, the most fundamental is proper time.

Introduced by Minkowski \cite{Minkowski}, proper time is the time measured by a clock moving along a given worldline. It represents the duration physically experienced by an observer following that trajectory and is independent of the coordinate system used to describe spacetime. Proper time thus occupies a privileged position in the theory, both physically and conceptually.

In this work, we propose an alternative route to incorporating gravitation into Special Relativity—one grounded entirely in physical principles rather than in the \textit{a priori} postulate of spacetime curvature. The construction begins with Galilean free fall, reflecting the universality of inertial response, and proceeds by extending Fermat’s principle \cite{Fermat}, which {tells} us that 
 light follows paths of extremal
travel time, 
to massive particles. Freely falling bodies are required to follow trajectories that extremize proper time. Lorentz invariance then ensures consistency with relativistic kinematics, while the retarded Newtonian potential introduced by Max Abraham in 1911 \cite{Abraham} provides the necessary dynamical input for the gravitational field.

When matter supports pressure, additional subtleties arise. Concepts such as active gravitational mass density and effective inertial mass density must be incorporated, blending insights from Newtonian gravity and Special Relativity to account for the dynamical role of pressure. With these elements in place, the invariant interval $ds^2$ emerges naturally in the case of a static gravitational field. By applying Lorentz transformations, the formulation is extended to moving matter, recovering—within the weak-field regime relevant to the Solar System— {exactly} the same solutions obtained from linearized General Relativity in harmonic gauge.

{

The conceptual significance of this approach is twofold. First, it places proper time, rather than curvature or coordinates, at the heart of the gravitational interaction, so that gravitation appears as a dynamical modulation of the flow of time itself. Second, when this reasoning is extended to stronger fields, the structure of the Ricci tensor in harmonic coordinates—together with the uniqueness results implied by Lovelock’s theorem—indicates that any consistent nonlinear completion must necessarily take the form of Einstein’s full field equations.

It is worth noting that alternative but dynamically equivalent formulations of gravitation, such as teleparallel and tetrad-based (frame)  \cite{Maluf2013, AldrovandiPereira2013}, also emphasize the close interplay between inertial and gravitational effects through the choice of reference frame. Since the present construction reproduces exactly the weak-field limit of General Relativity in harmonic gauge, its predictions are fully consistent with those obtained in teleparallel gravity, which is known to be equivalent to General Relativity at the level of the field equations. However, the goal of this work is to remain within a metric-based formulation and to derive the weak-field structure directly from dynamical and equivalence-principle considerations. A detailed reformulation in terms of proper frames or tetrads is therefore left for future investigation.

Thus, although the geometrical formulation of General Relativity has proved extraordinarily successful—particularly in strong-field regimes such as black holes—the present work concentrates on reconstructing the weak-field structure of General Relativity directly from dynamical first principles. In this sense, the geometric theory emerges not as an independent postulate, but as the natural codification of a deeper dynamical framework already implicit in the weak-field regime.

}

\section{Special Relativity}

{ 

Einstein appeared on the scientific scene in 1905 with the publication of his paper
{\it On the Electrodynamics of Moving Bodies} \cite{Einstein1905},
which is the foundational text of Special Relativity.

The work is remarkably sober and purely operational: Einstein avoids mechanical models—such as the classical ether—and restricts himself to analyzing what clocks and rulers actually measure when observers move uniformly with respect to one another.

The starting point consists of two postulates:
\begin{enumerate}
\item {\bf Principle of Relativity.} Einstein starts from the principle formulated by Galileo in his {\it Discorsi} \cite{Galileu}: on Earth, the laws of mechanics are the same for all observers who move with constant velocity.
\begin{quote}
{\it If a man were inside a ship sailing uniformly and without jolts, and he could neither see the sea nor feel the wind or the air, he would not, by any experiment performed inside the ship, be able to tell whether he is at rest or in motion.}
\end{quote}

What Einstein does is to take an enormous conceptual step: he considers Galilean relativity in {\it inertial} systems—those in which free test particles move with uniform rectilinear motion—and elevates this local principle of mechanics to a universal one, asserting that all physical laws, including those of electrodynamics and optics, must have the same form in all systems moving with constant velocity.

\item {\bf Constancy of the speed of light.} Light propagates in vacuum with a constant speed, independently of the motion of the emitting source.

Note that Maxwell’s equations show that the electromagnetic fields generated by a charge or a current propagate in vacuum with a fixed speed—in our units $c=1$. This propagation, with a delay determined by distance, is independent of the motion of the emitting source. Thus, the speed of light does not depend on who generated it: it is an intrinsic property of the vacuum, establishing a universal limit that underlies the constancy of the speed of light in all inertial frames.
\end{enumerate}

These two postulates, dispensing with any mechanical medium such as the ether, lead almost inevitably to the conclusion that the transformations relating two observers in uniform relative motion cannot be the Galilean ones, but must instead be the Lorentz transformations. Einstein does not start from Maxwell’s equations to obtain the Lorentz transformations; he arrives at Lorentz directly from the mutual compatibility of his two postulates.

To this end, Einstein presents a technically elaborate analysis of the propagation of light in two inertial systems with coordinates $(t,x,y,z)$ and $(\tau,\xi,\eta,\zeta)$, in uniform relative motion. This analysis involves the introduction of a third inertial system, likewise in uniform translational motion, and the coordinated use of clocks in all three systems to obtain the Lorentz transformations. Remarkably, however, in a footnote he points out:

\begin{quote}
    {\it The equation of the Lorentz transformation may be more simply deduced directly from the condition that in virtue of those equations the relation $x^2+y^2+z^2=t^2$ shall have as its consequence the second relation $\xi^2+\eta^2+\zeta^2=\tau^2$.}
\end{quote}

Once these transformations are established, Einstein applies his operational method to electromagnetism. He demands that Maxwell’s equations retain the same form in all inertial frames and, using this invariance criterion, deduces the transformation laws of the electric and magnetic fields. The result is extraordinary: magnetism emerges as a relativistic effect of an electric field as seen from another frame of reference.

In 1905 Einstein thus {constructed} a theory of space and time completely freed from supplementary mechanical hypotheses: it is a theory formulated strictly in terms of clocks, rulers, observers, and physical principles, and this conceptual economy was one of the keys to his revolution.

\

Later, in 1908, Hermann Minkowski entered the scene with his famous article {\it Raum und Zeit} \cite{Minkowski}, in which he gave definitive mathematical form to Einstein’s Special Relativity. While Einstein’s 1905 text still retained a certain literary tone—which, despite its conceptual value, did not yet anticipate its more mature formulation—Minkowski introduced a rigorous geometric structure that would become fundamental. It is in this article that the key element of special relativity appears for the first time: the spacetime invariant $ds^2$ and the concept of proper time.

\

{
To deduce the form of the proper time, we consider a test particle moving with velocity ${\bf v}$ with respect to an inertial system $\Sigma$ with coordinates $(t,{\bf r})$ where we have introduced the notation ${\bf r}=(x,y,z)$. We will denote by $ds$  the differential of time of the clock that moves with the test particle (the differential of proper time). The  relationship between the coordinate time $t$ --the proper time of the observer at rest with respect {\color{red}to} $\Sigma$-- and the proper time of the particle, must be of the form 
\begin{eqnarray}
ds = F(\mathbf v)\,dt \qquad \mbox{with} \qquad F({\bf 0})=1.
\end{eqnarray}

Let us consider another inertial system  $\Sigma'$ moving with respect to $\Sigma$ with 
velocity  ${\bf u}=(u,0,0)$ along the  $OX$ axis.

\medskip
The Lorentz transformation gives:
\begin{eqnarray}
dt' = \gamma_u\!\left(dt - {u\,dx}\right)
     = \gamma_u\!\left(1 - {u\,v_x}\right) dt,
\end{eqnarray}
where $\gamma_u = (1-u^2)^{-1/2}$ 
is the 
 Lorentz factor and
 we have used that $v_x=\frac{dx}{dt}$, being 
 $v_x$  the  component of the velocity ${\bf v}$ 
 in the direction of the 
 $OX$ axis.

\medskip

On the other hand,
from the formula of the composition of velocities we have
\begin{eqnarray}
v'_x=\frac{v_x-u}{1-{u\,v_x}},\qquad
\mathbf v'_\perp=\frac{\mathbf v_\perp}{\gamma_u\!\left(1-{u\,v_x}\right)},
\end{eqnarray}
where, to simplify,
we have used the notation
 ${\bf v}_{\perp}=(v_y,v_z)$.

\medskip

{Since the proper time is an invariant —i.e., it has the same value when measured in both systems— we have:}
\begin{eqnarray}
F(\mathbf v')\,dt' = F(\mathbf v)\,dt,
\end{eqnarray}
and replacing $dt'$:
\begin{eqnarray}\label{eq:funcional}
F(\mathbf v')\,\gamma_u\!\left(1 - {u\,v_x}\right)
= F(\mathbf v).
\end{eqnarray}

This functional relation must be true for any
 $\mathbf v$ {and} any $\mathbf u$.

To find the explicit form of
 $F$,  we consider the test particle at rest in the system $\Sigma'$, that is $\mathbf v'=\mathbf 0$, and {therefore its} velocity in  $\Sigma$ is $\mathbf v=\mathbf u$.
Replacing in  \eqref{eq:funcional}:
\begin{eqnarray}
F(\mathbf 0)\,\gamma_u\!\left(1-{u^2}\right)=F(\mathbf u).
\end{eqnarray}

Since $1-u^2=\gamma_u^{-2}$, we get
\begin{eqnarray}
F(\mathbf u)= F(\mathbf 0)\,\frac{1}{\gamma_u}=\frac{1}{\gamma_u}  
=\sqrt{1-{u^2}}.
\end{eqnarray}

And taking into account that $\mathbf u$
is an arbitrary velocity, we can write
\begin{eqnarray}
F(\mathbf v)=\sqrt{1-|\mathbf v|^2} \Longrightarrow ds = \sqrt{1-|\mathbf v|^2}\,dt,
\end{eqnarray}
and writing  ${\bf v}=\frac{d{\bf r}}{dt}$, 
the square of the differential of proper time --the Minkowski invariant-- takes the familiar form in the system $\Sigma$:
\begin{eqnarray}\label{Minkowskiinvariant0}
ds^2 =  dt^2 - d\mathbf r\cdot d\mathbf r\equiv \eta(d{\bf x},d{\bf x}),
\end{eqnarray}
where $d\mathbf r\cdot d\mathbf r=dx^2+dy^2+dz^2
$ is the Euclidean inner product and ${\bf x}\equiv (t,\bf {r})$.

In this way, the proper time elapsed along the worldline of a particle moving with constant velocity is simply
\begin{eqnarray}s=\int ds.\end{eqnarray}

The same reasoning extends to a particle in arbitrary motion. Its trajectory can be approximated by a polygonal path composed of infinitesimal segments, along each of which the velocity is nearly constant. The total proper time is obtained by summing the contributions of each segment and then taking the limit as the segment lengths go to zero.

This procedure leads to the general expression of the proper time elapsed between two events 
$P_1=(t_1,{\bf r}_1)$ and  $P_2=(t_2,{\bf r}_2)$:
\begin{eqnarray}
s \;=\; \int {ds}
\;=\;  \int_{t_1}^{t_2}
\sqrt{\,1 - |{\bf v}(t)|^2\,}\; dt,
\end{eqnarray}
where ${\bf v}(t)$
is the instantaneous velocity along the trajectory.

Thus, Minkowski introduced proper time as the intrinsic time measured by a clock moving with the particle: invariant, independent of the reference frame, and reflecting an intrinsic property of the particle’s motion in spacetime.
{This invariant will play the central dynamical role in the developments that follow.}

}

\medskip

}

\section{ The Equivalence Principle}

In 1907, Einstein wrote a synthesis article on special relativity entitled \textit{On the Relativity Principle and the Conclusions Drawn from It} \cite{Einstein1907}. At the end of this work, Einstein introduced for the first time the Equivalence Principle. In this first formulation one can read:
\begin{quote}
{\it Let us consider two reference systems $\Sigma_1$ and $\Sigma_2$. Suppose that $\Sigma_1$ is accelerated in the direction of the $OX$ axis,
and that $a$ is its acceleration, constant in time. Suppose that $\Sigma_2$ is at rest, but situated in a homogeneous gravitational field that imparts to all objects an acceleration $-a$ in the direction of the $OX$ axis.

As far as we know, the laws of physics referred to
$\Sigma_1$ do not differ from those referred to $\Sigma_2$;
this follows from the fact that all bodies are accelerated equally in this gravitational field.
Thus, according to our experience, we have no reason to suppose that the systems
$\Sigma_1$ and $\Sigma_2$
differ in any way, and therefore we shall henceforth assume the complete physical equivalence between a homogeneous gravitational field and the corresponding acceleration of the reference system.}
\end{quote}

A notable feature of this first version of the Equivalence Principle is its global character. Einstein identifies here a fundamental fact: for him, absolute acceleration with respect to Newtonian absolute space now loses physical content, just as absolute velocity has no meaning in Special Relativity. This allows him to extend the principle of relativity to uniformly accelerated motions and, at the same time, thanks to their equivalence with uniform gravitational fields, to include such fields within Special Relativity. The decisive step will be to extend the principle of relativity to any arbitrarily moving system and, in this way, to incorporate gravitation itself into Special Relativity.

This dynamical interpretation of the Equivalence Principle stands in marked contrast with the geometric viewpoint that became dominant following Pauli’s reformulation. It is therefore essential to distinguish clearly between Einstein’s original understanding of the Equivalence Principle, developed between 1907 and 1911, and the later local reinterpretation introduced by Pauli within the fully developed framework of General Relativity.

In Einstein’s early formulation, the Equivalence Principle expressed a global physical correspondence between a uniformly accelerated reference frame and a homogeneous gravitational field. In line with the classical ideas of D’Alembert \cite{Dalembert}, it embodied a true dynamical equivalence between inertial and gravitational forces: the inertial reaction force $-ma$
 precisely counteracts the gravitational force when inertial and gravitational masses coincide. From this standpoint, the experience of weightlessness in free fall arises from the exact cancellation of two real and opposing forces acting on the same body, rather than from any geometrical disappearance of gravity.

By contrast, when dealing with general gravitational fields, the formulation introduced by Pauli in 1921 adopts a fundamentally different, local notion of equivalence. In his treatment \cite{Pauli1921}, Pauli argued that:
\begin{quote}
\textit{
“For every infinitely small world region (i.e., a region so small that the space- and time-variation of gravity can be neglected) there always exists a coordinate system … in which gravitation has no influence either on the motion of particles or on any other physical process.”}
\end{quote}


This reformulation redirected the focus from the global physical behavior of gravitational systems to the local geometrical properties of spacetime.
The Equivalence Principle ceased to express a physical equivalence between acceleration and gravitation and was instead recast as a mathematical condition on the local structure of the metric.
At any spacetime point, one may always choose a coordinate system in which the metric reduces locally to its Minkowski form and the associated Christoffel symbols vanish.
This situation is directly analogous to the introduction of geodesic normal coordinates on a curved Gaussian surface, where, within an infinitesimal neighborhood, geodesics appear as straight lines and the effects of curvature become undetectable.
In both contexts, the apparent disappearance of curvature is not a physical effect but a consequence of a coordinate choice that removes it at first order.
In this way, Pauli’s formulation recast Einstein’s original principle—initially grounded in the dynamical relationship between inertia and gravitation—into a purely geometrical statement concerning the local flatness of spacetime.
As a result, the deep physical insight of Einstein’s early view, in which gravitation and inertia represented two facets of a single dynamical phenomenon, was replaced by a purely kinematical property of the affine connection.

Although this local formulation plays a central role in the mathematical foundations of General Relativity, it differs substantially from Einstein’s original conception developed between 1907 and 1911, which was physical and global rather than infinitesimal and geometrical.
Einstein’s original statement described a genuine physical equivalence between a homogeneous gravitational field and a uniformly accelerated reference frame, valid over a finite spacetime domain.
By contrast, Pauli’s version merely ensures that, in an infinitesimal neighborhood of any spacetime point, the motion of freely falling particles in an arbitrary gravitational field can be locally approximated as uniform and rectilinear through an appropriate coordinate transformation.

The approach developed in this work returns to Einstein’s original physical insight by interpreting the Equivalence Principle as a dynamical law linking inertia and gravitation.
From this perspective, the gravitational modification of the spacetime invariant $ds^2$
 is not introduced as a geometric postulate, but instead arises dynamically from the balance between inertial and gravitational forces, thereby preserving the physical content of Einstein’s early conception within a fully relativistic and self-consistent framework.

\subsection{Application of the Equivalence Principle}

As we shall see, for Einstein the Equivalence Principle plays a very important practical role. This principle states that everything that happens inside a uniformly accelerated system can also be applied to a uniform gravitational field. This allows him to use the behavior of an accelerated system to infer how gravitational effects should manifest themselves in more general and complex situations.

Three years and some months after his 1907 article, Einstein published his second work devoted to gravitation, {\it On the Influence of Gravitation on the Propagation of Light} \cite{Einstein1911}. In this paper he takes a decisive conceptual step: he shows that the presence of a gravitational field alters the passage of time. His reasoning combines Special Relativity with the Equivalence Principle and constitutes the first explicit formulation of what we now call gravitational redshift.

In order to present this idea in an even more transparent way, we shall provide here a direct derivation based on the non-relativistic Doppler effect.

Thus, let us consider the classical Doppler effect formula:
\begin{eqnarray}
    \nu = \left(\frac{v + v_r}{v + v_s}\right)\nu_0,
\end{eqnarray}
where $\nu$ is the observed frequency, $\nu_0$ the emitted frequency, $v$ the wave speed, $v_r$ the velocity of the receiver, and $v_s$ the velocity of the source.

For light ($v \equiv 1$) and with the source at rest, this expression reduces to
\begin{eqnarray}
    \nu = \left(1 + v_r\right) \nu_0.
\end{eqnarray}

If the receiver has a constant acceleration $a$ in the $OZ$ direction, its velocity at time $t$ will be
$v_r = a t = a h$, where $h$ is the height at which the receiver is located. We thus obtain
\begin{eqnarray}
    \nu = \left(1 + a h\right) \nu_0.
\end{eqnarray}

Applying the Equivalence Principle allows one to replace $a h$ by the homogeneous gravitational potential $\Phi$, yielding the approximate result for the gravitational frequency shift:
\begin{eqnarray}
    \nu = \left(1 + \Phi\right) \nu_0.
\end{eqnarray}

This derivation is strictly non-relativistic and valid only for small velocities and weak fields. Nevertheless, it shows in a very direct way the connection between acceleration, gravitation, and frequency shift. Showing that gravity alters the flow of time, which as we will see in {the} next Section,  is the key point to extend the Special Relativity to include gravity.

{\section{
Gravity Without Geometry}}

This section is based {on} three recent papers in collaboration with Emilio Elizalde
\cite{HE2025,HE2025a,HE2026}.

Our aim is to extend the framework of Special Relativity to include gravitational effects.
To do {this},  we have to realize, as we have already seen in {the} previous section,  that gravity modifies the {passage} of time, {which} means that 
the Minkowski invariant --the {differential} of proper time-- has to be modified by one that includes the gravitational potential.
{Our guiding requirement in determining its explicit form is that it must reproduce Galilean free fall in a uniform gravitational field $\Phi(z)=gz$, based on the following extension of Fermat’s principle: {\it A test particle moving under the action of gravity follows a trajectory that extremizes its proper time.}
}

{

{We consider} the isotropic invariant:
\begin{eqnarray}\label{uniform}
    ds^2 = A(z) dt^2 - B(z) d{\bf r} \cdot d{\bf r},
\end{eqnarray}
{which leads to the normalization of the four-velocity}
\begin{eqnarray}\label{fourvelocity}
     1=L\equiv A(z)\left(\frac{dt}{ds}\right)^2-B(z) \left|\frac{d{\bf r}}{ds}\right|^2.
\end{eqnarray}

Minimizing the proper time,
\begin{eqnarray}
\delta \int ds = 0,
\end{eqnarray}
for a test particle initially at rest and freely falling along the {$z$-axis},
{which, since $L=1$, is equivalent to minimizing \cite{Landau}}
\begin{eqnarray}
    \int L\, ds,
\end{eqnarray}
together with (\ref{fourvelocity}), yields, via the Euler–Lagrange equations,
{
\begin{eqnarray}
    \frac{d}{ds}\left(\frac{\partial L}{\partial (dz/ds)}\right)
    =\frac{\partial L}{\partial z},
\end{eqnarray}
}
the corresponding dynamical equation:
\begin{align}
  m \frac{d^2 z}{ds^2} = - \frac{m}{2 A(z) B(z)} \left[
    \partial_z A + \partial_z (A B) \left( \frac{dz}{ds} \right)^2 \right].
\end{align}

}

To obtain the free fall law, it is necessary to impose that the velocity-dependent term on the right-hand side vanish. This condition holds when
$
{A(z) B(z) = C},
$
where ${C}$ is a constant that can be taken as unity by rescaling coordinates. 

Thus, we obtain:
\begin{eqnarray}
  - m \frac{d^2 z}{ds^2} - \frac{m}{2} \partial_z A = 0,
\end{eqnarray}
and to obtain the Galilean free fall we have to demand 
\begin{eqnarray}\label{exact}
a_{\rm p} = \frac{d^2 z}{ds^2}, \qquad 
A(z) = b + 2 \Phi(z), \qquad B(z) = \frac{1}{b + 2\Phi(z)},
\end{eqnarray}
where ${b}$ is a constant that must equal unity to recover the Minkowski invariant when the field vanishes. Thus, the invariant takes the form
\begin{eqnarray}\label{uniforme}
    ds^2=(1+2gz)\,dt^2-(1+2gz)^{-1}\,d{\bf r}\cdot d{\bf r}.
\end{eqnarray}

We conclude, therefore, that for a uniform gravitational field described by
$\Phi(z)=gz$, the trajectory of a freely falling particle obeys exactly the
Galilean law when the acceleration is identified with the spatial
component of the four--acceleration,
\begin{equation}
\frac{d^2 z}{ds^2}=-g .
\end{equation}
This result holds without any restriction on the particle velocity and does
not rely on a weak--field approximation.

\medskip

\begin{remark}
If one instead adopts, as Einstein did in 1912 \cite{Einstein1912a,Einstein1912b},
the invariant
\begin{eqnarray}\label{A}
    ds^2=(1+2gz)\,dt^2-d{\bf r}\cdot d{\bf r},
\end{eqnarray}
the corresponding equation of motion reads
\begin{eqnarray}
    \frac{d^2z}{ds^2}
    =
    -\frac{g}{1+2gz}
    \left(1+\left(\frac{dz}{ds}\right)^2\right).
\end{eqnarray}
In this case, the Galilean law is recovered only in the simultaneous limits of
weak gravitational field, $|gz|\ll 1$, and low velocities,
$\left|\frac{dz}{ds}\right|\ll 1$.

By contrast, linearizing the invariant (\ref{uniforme}) leads to
\begin{eqnarray}\label{B}
    ds^2=(1+2gz)\,dt^2-(1-2gz)\,d{\bf r}\cdot d{\bf r},
\end{eqnarray}
from which one finds
\begin{eqnarray}
    \frac{d^2z}{ds^2}
    =
    -g\,
    \frac{1-4gz\left(\frac{dz}{ds}\right)^2}{1-4g^2z^2}
    \simeq
    -g ,
\end{eqnarray}
whenever $|gz|\ll 1$, independently of the particle velocity.

This provides a clear dynamical criterion: the invariant (\ref{B}) preserves the
Galilean law in the weak--field regime without imposing any restriction on the
velocity, whereas the invariant (\ref{A}) does not. For this reason, (\ref{B})
constitutes a physically consistent weak--field limit of (\ref{uniforme}).

An additional—and conceptually crucial—difference concerns the harmonic gauge.
The invariant (\ref{A}) does not satisfy the linear harmonic gauge condition.
This fact sheds light on why, in the \emph{Entwurf} theory \cite{Grossmann},
Einstein and Grossmann ultimately rejected the Ricci tensor in 1913: only within
the linearized harmonic gauge does the Ricci tensor reduce, in the weak--field
limit, to the classical Poisson equation.

By contrast, the invariant (\ref{B}) satisfies the harmonic gauge already at the
linear level. As a result, it naturally leads to the correct Newtonian limit and
allows for a consistent nonlinear completion based on the Ricci tensor.
\end{remark}

\subsection{Static Fields}

{
Building on the previous result for a uniform gravitational field, 
we consider, for a static, pressureless configuration, the following invariant \cite{Synge}:
\begin{eqnarray}\label{conformastat}
   ds^2=(1+2\Phi({\bf r}))\,dt^2-(1+2\Phi({\bf r}))^{-1} d{\bf r}\cdot d{\bf r},
\end{eqnarray}
{which now leads to the following normalization of the four-velocity}
\begin{eqnarray}
    1=L\equiv (1+2\Phi({\bf r}))\,\dot{t}^2-(1+2\Phi({\bf r}))^{-1} \dot{\bf r}\cdot \dot{\bf r},
\end{eqnarray}
{where the dot denotes differentiation with respect to the proper time.}

{Minimizing the proper time,
\begin{eqnarray}
\int ds = \int L\,ds,
\end{eqnarray}
together with the constraint $L=1$},
the Euler–Lagrange equations yield:}
\begin{eqnarray}\label{geodesicequation}
   \ddot{\bf r}
    =-\nabla\Phi
    -2
    \frac{|\dot{\bf r}|^2}{1+2\Phi}\nabla^{\perp}\Phi ,
\end{eqnarray}
where  
\begin{eqnarray}
\nabla^{\perp}\Phi=
\nabla\Phi-\frac{\dot{\bf r}\cdot \nabla\Phi}{|\dot{\bf r}|^2}\dot{\bf r}
\end{eqnarray}
is the orthogonal projection of $\nabla\Phi$ onto the subspace orthogonal to $\dot{\bf r}$.

\

Therefore, 
when the velocity is parallel to the gradient of $\Phi$, {the exact Newtonian equation is recovered:} 
\begin{eqnarray}
   \ddot{\bf r}=-\nabla\Phi.
\end{eqnarray}

As an example, consider a freely falling body near the Earth’s surface moving along a purely radial trajectory,  
${\bf r}(t)=r(t){\bf r}$.  
In this case, the exact equation of motion reduces to
\begin{eqnarray}
   \ddot{r}=-\frac{MG}{r^2}\cong 9.8~{\rm m/s^2},
\end{eqnarray}
in {SI} units.   

\

Moreover, this invariant naturally implies the conservation of energy. 
Indeed, taking the inner product of equation \eqref{geodesicequation}, with the momentum ${m\dot{\bf r}}$, yields
\begin{eqnarray}
   \frac{m|\dot{\bf r}|^2}{2}+m\Phi({\bf r})=E.
\end{eqnarray}

\

Another compelling feature of this metric is that it predicts the same deflection of light as the {Schwarzschild} solution, namely \cite{Weinberg}
\begin{eqnarray}
   \delta\phi=\frac{4MG}{b},
\end{eqnarray}
where $b$ denotes the impact parameter, and also the precession of the perihelion of Mercury.

\subsection{Moving masses}\label{movingmasses}

We now study, in the weak limit,  the gravitational effect produced in the Minkowski invariant 
$ds^2$ when the configuration is not static, that is, when the sources are in motion.  

\begin{remark}
Here we restrict ourselves to the weak--field regime, retaining only the
terms linear in the gravitational potential. The reason is purely physical.
In order to extend the static invariant $ds^2$ to moving matter, we apply
Lorentz transformations, which are defined in the absence of gravitation and
therefore encode gravity only at zeroth order.

As a consequence, a Lorentz boost can be consistently applied only to the
linear part of the gravitational invariant. Higher--order terms in the
potential represent genuine gravitational self--interaction effects, which
are not captured by a kinematical transformation defined on flat spacetime.
Keeping such terms would therefore mix different orders of approximation in
an inconsistent way.

For this reason, only the linearized form of the static invariant can be
meaningfully promoted to the case of moving sources by means of Lorentz
transformations. The unique consistent extension of this construction beyond
the weak--field regime will be addressed in the final section.
\end{remark}

{

Thus, in a frame, namely $\Sigma$, where the masses are at rest, in the weak-field limit we have
\begin{eqnarray}\label{conformastat_weak}
   ds^2=(1+2\Phi({\bf r}))\,dt^2-(1-2\Phi({\bf r}))\, d{\bf r}\cdot d{\bf r},
\end{eqnarray}
{where the potential $\Phi$ satisfies the Poisson equation $\Delta\Phi=4\pi G\rho$.
 Note that, under a Lorentz boost with velocity ${\bf v}$, the static Poisson equation generalizes naturally to the wave equation $\Box \Phi = -4\pi G \rho$, where
 $\Box = \partial_t^2 - \Delta$ is the d’Alembert operator, 
 incorporating retardation effects, while the linearized invariant transforms consistently.}

\

Following Kenneth Nordtvedt \cite{Kenneth}, let us consider another frame $\Sigma'$ moving with three-dimensional velocity $-{\bf v}'$ with respect to the frame $\Sigma$.  
Equivalently, $\Sigma$ moves with velocity
\begin{eqnarray}
   {\bf v}'=\frac{d{\bf r}'}{dt'}
\end{eqnarray}
with respect to $\Sigma'$.

\

Let $\gamma$ be the Lorentz factor
\begin{eqnarray}
   \gamma=\frac{1}{\sqrt{1-|{\bf v}'|^2}} .
\end{eqnarray}

{
Then, the {Lorentz transformation that leaves the Minkowski interval invariant,
i.e., $ds^2=dt^2-d{\bf r}\cdot d{\bf r}=dt'^2-d{\bf r}'\cdot d{\bf r}'$, reads}
{\begin{eqnarray}
   {\bf r}=\gamma\left[
    \frac{1}{\gamma}{\bf r}'+\frac{\gamma}{1+\gamma}({\bf v}'\cdot {\bf r}'){\bf v}'
    -{\bf v}'t'
    \right],\qquad 
    t=\gamma(t'-{\bf v}'\cdot {\bf r}').
\end{eqnarray}}
}

\

The key point is that, given two constants $a$ and $b$, the quadratic form
$a\,dt^2 + b\, d{\bf r}\cdot d{\bf r}$ transforms into
\begin{eqnarray}
\gamma^2(a+b|{\bf v}'|^2)\, dt'^2
-2\gamma^2(a+b)({\bf v}'\cdot d{\bf r}')\, dt'
+ b\, d{\bf r}'\cdot d{\bf r}'
+ \gamma^2(a+b)({\bf v}'\cdot d{\bf r}')^2.
\end{eqnarray}

{Then, removing the primes, the invariant in $\Sigma'$, {generated by a source moving with velocity ${\bf v}$}, can be written as
\begin{eqnarray}\label{LORENTZ}
ds^2 &=&
(1 - 2\Phi + 4\Upsilon)\, dt^2
- 8\, {\mathbf{N}} \cdot d\mathbf{r}\, dt
- (1 - 2\Phi)\, d\mathbf{r} \cdot d\mathbf{r}
+ 4\, \frak{u}(d{\bf r}, d{\bf r}),
\end{eqnarray}
where the new potentials are functions of the potential $\Phi$, {which, after the Lorentz transformation, must satisfy Abraham's equation \cite{Abraham,Nordstrom}}
\begin{eqnarray}
\Phi(t,{\bf r}) = -G\int \frac{\rho(t_{\rm r},\bar{\bf r})}{|{\bf r}-\bar{\bf r}|}\, d\bar{V}
\Longrightarrow \Box \Phi = -4\pi G \rho,
\end{eqnarray}
}
where $t_{\rm r} = t - |{\bf r}-\bar{\bf r}|$ is the retarded time.
{The velocity ${\bf v}$ enters through
\begin{eqnarray}
    \Upsilon = \frac{\Phi}{1-|{\bf v}|^2}, \qquad
    {\bf N} = \frac{\Phi}{1-|{\bf v}|^2} {\bf v}, \qquad
    \frak{u}_{ij} = \frac{\Phi}{1-|{\bf v}|^2} v_i v_j.
\end{eqnarray}

For small velocities, keeping only linear terms in ${\bf v}$, these reduce to
\begin{eqnarray}
    \Upsilon \simeq \Phi, \qquad
    {\bf N} \simeq \Phi {\bf v}, \qquad
    \frak{u}_{ij} \simeq 0,
\end{eqnarray}}
and the invariant takes the simplified form
\begin{eqnarray}\label{LORENTZ_linear}
ds^2 \simeq
(1 + 2\Phi)\, dt^2
- 8\, {\mathbf{N}} \cdot d\mathbf{r}\, dt
- (1 - 2\Phi)\, d\mathbf{r} \cdot d\mathbf{r}.
\end{eqnarray}

\

To construct the potentials produced by a continuous distribution of matter whose elements may move with different velocities, we proceed as follows. The contributions from infinitesimal volume elements combine linearly, each element contributing independently to the invariant. Each element moves with its local velocity ${\bf v}_k$, and the corresponding Lorentz transformation is applied separately. The full invariant is then obtained by superposing all these contributions, which formally amounts to integrating over the entire distribution.
{This linear superposition is justified in the weak-field regime considered here, where the metric deviations from Minkowski spacetime are small and the gravitational potentials enter only at first order, consistently with the standard linearized treatment of General Relativity.

}

Specifically, dividing the matter into $N$ small cells of density $\rho_k$ and volume $\Delta \bar{V}_k$, the invariant \eqref{conformastat_weak} can be approximated as
\begin{eqnarray}
ds^2 \simeq \sum_{k=1}^N
\Bigg[
\Big(\frac{1}{N} + 2\Phi_k \Big) dt^2
- \Big( \frac{1}{N} - 2\Phi_k \Big) d{\bf r}\cdot d{\bf r}
\Bigg],
\end{eqnarray}
where
\begin{eqnarray}
\Phi_k({\bf r}) \equiv -G\, \frac{\rho_k \Delta \bar{V}_k}{|{\bf r}-\bar{\bf r}_k|}.
\end{eqnarray}

Applying a Lorentz transformation with velocity ${\bf v}_k$ to the $k$-th cell, the invariant becomes
\begin{eqnarray}
ds^2 \simeq \sum_{k=1}^N
\Bigg[
\Big(\frac{1}{N} - 2\Phi_k + 4 \Upsilon_k \Big) dt^2
- 8 {\bf N}_k \cdot d{\bf r}\, dt
- \Big(\frac{1}{N} - 2\Phi_k \Big) d{\bf r}\cdot d{\bf r}
+ 4 (\frak{u}_k)_{ij} dx^i dx^j
\Bigg],
\end{eqnarray}
with
\begin{eqnarray}
\Upsilon_k = \frac{\Phi_k}{1-|{\bf v}_k|^2}, \quad
{\bf N}_k = \frac{\Phi_k}{1-|{\bf v}_k|^2} {\bf v}_k, \quad
(\frak{u}_k)_{ij} = \frac{\Phi_k}{1-|{\bf v}_k|^2} ({\bf v}_k)_i ({\bf v}_k)_j,
\end{eqnarray}
where $\Phi_k$ is now the retarded potential of the $k$-th element after the Lorentz transformation.

Finally, taking the continuum limit yields the invariant \eqref{LORENTZ}, with the potentials satisfying
\begin{eqnarray}
\Box \Upsilon = -\frac{4\pi G \rho}{1-|{\bf v}|^2}, \quad
\Box {\bf N} = -\frac{4\pi G \rho}{1-|{\bf v}|^2} {\bf v}, \quad
\Box \mathfrak{u} = -\frac{4\pi G \rho}{1-|{\bf v}|^2} {\bf v}^{\flat} \otimes {\bf v}^{\flat}.
\end{eqnarray}
Here, the ``flat'' operator denotes the dual one-form associated with the vector field ${\bf v}$.

}

\

{
Therefore, expression~\eqref{LORENTZ} coincides with the result obtained in linearized General Relativity in the harmonic gauge,
when the stress-tensor is considered in Minkowski space, and thus, 
 the four-velocity is given by
\begin{eqnarray}
{\bf u}=\frac{1}{\sqrt{1-|{\bf v}|^2}}(1, {\bf v}).
\end{eqnarray}

{

}

{

{

To verify the consistency of the previous results with standard linearized General Relativity, we proceed as follows.

Writing the invariant in coordinates as 
\[
ds^2 = g_{\mu\nu} dx^{\mu} dx^{\nu},
\]
with $g_{\mu\nu} = \eta_{\mu\nu} + h_{\mu\nu}$, where $\eta_{\mu\nu}$ is the Minkowski metric and $h_{\mu\nu}$ is a small perturbation, the linearized Ricci tensor in harmonic coordinates takes the form
\begin{eqnarray}
R_{\mu\nu} = -\frac{1}{2}\Box h_{\mu\nu}.
\end{eqnarray}

{

Consequently, the linearized form of Einstein’s field equations \cite{Einstein1916},
\begin{eqnarray}\label{EFE}
R_{\mu\nu} = 8\pi G \left(T_{\mu\nu} - \frac{1}{2} g_{\mu\nu} T \right),
\end{eqnarray}
where for a pressureless fluid the stress–energy tensor is
\begin{eqnarray}
T_{\mu\nu} = \rho\, u_{\mu} u_{\nu}, 
\qquad 
T = \rho,
\end{eqnarray}
takes the form
\begin{eqnarray}
\Box h_{\mu\nu}
=
-16\pi G \left(T_{\mu\nu} - \frac{1}{2} \eta_{\mu\nu} T \right)
\quad\Longleftrightarrow\quad
\Box\!\left(h_{\mu\nu}-\frac{1}{2}\eta_{\mu\nu}h\right)
=
-16\pi G\,T_{\mu\nu},
\end{eqnarray}
where we have used that $R=-8\pi G T$, obtained by taking the trace of Einstein’s field equations~(\ref{EFE}), together with the linearized relation in harmonic coordinates
\begin{equation}
R=-\frac{1}{2}\Box h,
\end{equation}
with $h \equiv \eta^{\mu\nu}h_{\mu\nu}$ the trace of $h_{\mu\nu}$.

}

\

This is precisely the form found in Eq.~(96a)–(96b) of \cite{Einstein1955} and in Eq.~(18.6)–(18.8) of \cite{MTW}. Recall that the stress-energy tensor, even when velocities are not assumed to be small, must be evaluated in Minkowski spacetime because $u_{\mu} = dx_{\mu}/ds$, and in the linear approximation
\[
ds^2 = \eta_{\mu\nu} dx^{\mu} dx^{\nu} = (1-|{\bf v}|^2)\,dt^2.
\]

As emphasized in \cite{MTW}:
\begin{quote}
{\it
It is important that the components of the stress-energy tensor $T^{\mu\nu}$, which appear in the linearized equations and in the conservation laws, are precisely the components one would calculate using Special Relativity (with $g_{\mu\nu} = \eta_{\mu\nu}$).
}
\end{quote}

Therefore, for $\mu=\nu=0$, we obtain
\begin{eqnarray}
\Box h_{00}
=
8\pi G \rho
-
16\pi G \frac{\rho}{1-|{\bf v}|^2},
\end{eqnarray}
which is satisfied for $h_{00}=-2\Phi+4\Upsilon$.

For $\mu=0$ and $\nu=j$, we have
\begin{eqnarray}
\Box h_{0j}
=
-16\pi G
\frac{v_j}{1-|{\bf v}|^2}\rho,
\end{eqnarray}
and thus $N_j=h_{0j}+h_{j0}=2h_{0j}$.

Finally, for $\mu=i$ and $\nu=j$,
\begin{eqnarray}
\Box h_{ij}
=
-8\pi G \delta_{ij}\rho
-
16\pi G
\frac{v_i v_j}{1-|{\bf v}|^2}\rho,
\end{eqnarray}
which leads to $h_{ij}=2\Phi\delta_{ij}+4\mathfrak{u}_{ij}$.

\medskip

Therefore, the metric potentials obtained through the present construction reproduce the structure predicted by linearized General Relativity in harmonic gauge. This correspondence emerges \emph{a posteriori} as a nontrivial consistency check of the framework, and is not imposed at any stage of the derivation.

}

}

\

This agreement shows that the formulation presented here reproduces, in the weak-field regime, the same physical predictions as linearized General Relativity in harmonic gauge, while dispensing with an explicit geometric description in terms of spacetime curvature. In this approach, the geometric features customarily attributed to spacetime arise instead as an effective representation of the fundamental dynamical interplay between matter and inertia.

It is also important to emphasize that, in the static case, the invariant (\ref{conformastat}) automatically fulfills the harmonic gauge condition. Since this condition remains invariant under Lorentz transformations, its boosted version—corresponding to moving sources—necessarily satisfies the same gauge requirement. This provides a natural explanation for why the expressions obtained for moving masses coincide exactly with those derived in linearized General Relativity formulated in harmonic coordinates.

As a consequence, the procedure guaranties that the resulting invariant consistently incorporates the combined contribution of all moving sources, leading to the same outcome as linearized General Relativity in the harmonic gauge, without requiring the full geometric machinery of the theory. Following Fock’s viewpoint \cite{Fock}, this result underscores the privileged physical role of harmonic coordinates, which he regarded as offering the most natural framework for describing the gravitational field. From this standpoint, the harmonic gauge should not be viewed merely as a technical convenience, but rather as a coordinate choice in which the dynamical content of gravity—encoded in the invariant—is expressed in its most direct and transparent form. This supports the idea that, at least in the weak-field limit, the essential physical content of General Relativity can be fully captured by an appropriately defined field-theoretic invariant, consistent with both the Equivalence Principle and Lorentz invariance, without recourse to the full Riemannian geometric framework.

{
\subsection{Matter supporting pressure}
\label{sec-C}

To go beyond the idealized case of pressureless matter, we now consider matter distributions capable of sustaining internal pressure. A natural and transparent framework for this purpose is provided by the Newtonian description of a homogeneous spherical body contracting or expanding in ordinary Euclidean space. Despite its simplicity, this setting already captures, in a consistent manner, the interplay between density, pressure, and the gravitational potential.

Let us consider a homogeneous sphere of time-dependent radius $a(t)R$, {where, as in cosmology , $a(t)$ is the {\it scale factor}}, and mass density $\bar{\sigma}(t)$. The total mass contained within the sphere is then
\begin{equation}
M = \frac{4\pi}{3}\,\bar{\sigma}(t)\,a^3(t)R^3.
\end{equation}

Choosing the center of the sphere as the origin, we parametrize the position of an arbitrary point $P$ by $a(t)\mathbf{r}$, so that its physical distance from the center is
{
\begin{eqnarray}
r(t) \equiv a(t)\,|\mathbf{r}|.
\end{eqnarray}
}
\medskip

By virtue of Newton’s shell theorem \cite{Newton} (see also p.~61 of \cite{Ryden}), only the mass enclosed within the radius $r$ contributes to the gravitational force at $P$. A test particle of mass $m$, momentarily at rest at this point, experiences a purely radial force given by
\begin{equation}
\mathbf{F} = -\frac{4\pi G m}{3}\,\bar{\sigma}(t)\,a(t)\mathbf{r}.
\end{equation}

Applying Newton’s second law, and exploiting the radial symmetry of the configuration, we obtain
{\begin{equation}
m\,\frac{d^2 r}{dt^2}
= -\frac{4\pi G m}{3}\,\bar{\sigma}\, r
\qquad\Longrightarrow\qquad
\frac{d^2 r}{dt^2}
= -\frac{G\,M(r)}{r^2}=-
\frac{G\,M\,r}{a^3R^3},
\end{equation}}
where
\[
M(r) = \frac{4\pi}{3}\, r^3 \bar{\sigma}
\]
denotes the mass enclosed within the sphere of radius $r$. {This coincides with the direct Newtonian derivation for a homogeneous expanding or contracting sphere \cite{Ryden,Mukhanov}.

}

\medskip

The same equation of motion may be derived from a variational principle. Consider the Newtonian Lagrangian
\begin{equation}
L = \frac{1}{2}\left(\frac{dr}{dt}\right)^2
+ \frac{4\pi G}{3}\, r^2 \bar{\sigma}
\qquad\Longleftrightarrow\qquad
L = \frac{1}{2}\left(\frac{dr}{dt}\right)^2 - V(r),
\end{equation}
with gravitational potential
\begin{equation}
V(r) = -\frac{G M(r)}{r}.
\end{equation}

{
Taking into account that $r=a|{\bf r}|$ the Lagrangian reads:
\begin{equation}\label{a_lagrangian}
L =|{\bf r}|^2\left[ \frac{1}{2}\left(\frac{da}{dt}\right)^2
+ \frac{4\pi G}{3}\, a^2 \bar{\sigma}\right]
\qquad\Longleftrightarrow\qquad
L =|{\bf r}|^2\left[ \frac{1}{2}\left(\frac{da}{dt}\right)^2 - \bar{V}(a)\right],
\end{equation}
with 
\begin{eqnarray}\label{pot}
\bar{V}(a)=-\frac{4\pi G}{3}a^2\bar{\sigma}.\end{eqnarray}

\

Performing the variation with respect the scale factor, the Euler-Lagrange equation reads:
\begin{equation}\label{ELnova}
|{\bf r}|^2\frac{d^2 a}{dt^2}
= \frac{4\pi G}{3}|{\bf r}|^2\,\frac{d}{da}(a^2\bar{\sigma})\quad
\Longleftrightarrow\quad
\frac{d^2 a}{dt^2}
= \frac{4\pi G}{3}\,\frac{d}{da}(a^2\bar{\sigma})
.
\end{equation}

}

}

\medskip

{

On the other hand, mass conservation within a comoving sphere of radius $r(t)$ implies
\begin{equation}
\frac{dM}{dt}=0,
\end{equation}
where
\begin{equation}
M(r)=\frac{4\pi}{3}r^3\bar{\sigma}=\frac{4\pi}{3}|{\bf r}|^3a^3\bar{\sigma}
.
\end{equation}

This expresses the Newtonian continuity equation for a homogeneous medium.

Therefore, the conservation law can be written as
\begin{equation}\label{a_conservation}
\frac{d}{dt}\!\left(a^3\bar{\sigma}\right)=0.
\end{equation}

Expanding the time derivative gives
\begin{equation}
3a^2\frac{da}{dt} \,\bar{\sigma}+
a^3\frac{d\bar{\sigma}}{dt}=0,
\end{equation}
which implies
\begin{equation}\label{sigmaderivada}
\frac{d\bar{\sigma}}{dt}=-\frac{3}{a}\frac{d a}{dt}\bar{\sigma}.
\end{equation}

To evaluate the derivative appearing in the Euler--Lagrange equation,
we note that the dynamical variable is the scale factor $a(t)$,
while $|{\bf r}|$ is a constant comoving coordinate.
Derivatives with respect to $a$ and $t$ are therefore related through the chain rule:
\begin{equation}
\frac{d}{da}=\frac{1}{da/dt }\frac{d}{dt}.
\end{equation}

Therefore,
\begin{equation}
\frac{d}{da}(a^2\bar{\sigma})
=\frac{1}{da/dt}\frac{d}{dt}(a^2\bar{\sigma}).
\end{equation}

Computing the time derivative yields
\begin{equation}
\frac{d}{dt}(a^2\bar{\sigma})
=2a\frac{da}{dt}\,\bar{\sigma}+a^2\frac{d\bar{\sigma}}{dt}.
\end{equation}

Substituting the relation for ${d\bar{\sigma}}/dt$ obtained in (\ref{sigmaderivada}),
we find
\begin{equation}
\frac{d}{dt}(a^2\bar{\sigma})
=2a\frac{da}{dt}\,\bar{\sigma}-3a\frac{da}{dt}\,\bar{\sigma}
=-a\frac{da}{dt}\,\bar{\sigma}.
\end{equation}

Consequently,
\begin{equation}
\frac{d}{da}(a^2\bar{\sigma})=-a\bar{\sigma}.
\end{equation}

Substituting this relation into the Euler--Lagrange equation (\ref{ELnova}), and taking into account once again that $r=a|{\bf r}|$,  yields
\begin{equation}
\frac{d^2 a}{dt^2}
= -\frac{4\pi G}{3}\,\bar{\sigma}\, a\quad \Longleftrightarrow
\quad
\frac{d^2 r}{dt^2}
= -\frac{4\pi G}{3}\,\bar{\sigma}\, r,
\end{equation}}
in agreement with the direct Newtonian derivation.
}

{

\medskip

\begin{remark}
An equivalent and more direct derivation can also be obtained as follows.
     From the conservation law (\ref{a_conservation}) we see that $\bar{\sigma}$ scales as $a^{-3}$, that is,
    \begin{eqnarray}
        \bar{\sigma}(t)=\bar{\sigma}_0\left(\frac{a_0}{a(t)}\right)^3,
    \end{eqnarray}
    and thus the Lagrangian (\ref{a_lagrangian}) becomes
\begin{eqnarray}
    L =|{\bf r}|^2\left[ \frac{1}{2}\left(\frac{da}{dt}\right)^2
+ \frac{4\pi G}{3}\, \frac{a_0^3\bar{\sigma}_0}{a}\right].\end{eqnarray}

Then, the Euler-Lagrange equation leads to:
\begin{eqnarray}
    \frac{d^2a}{dt^2}=\frac{4\pi G}{3}a_0^3\bar{\sigma}_0\frac{d}{da}\left(\frac{1}{a}\right)=-\frac{4\pi G}{3}\frac{a_0^3\bar{\sigma}_0}{a^2}=-\frac{4\pi G}{3}\bar{\sigma}a\quad\Longleftrightarrow\quad
    \frac{d^2r}{dt^2}=-\frac{4\pi G}{3}\bar{\sigma}r.
\end{eqnarray}

\end{remark}

\medskip

\medskip

We thus see that the variational formulation, together with the conservation of mass in a comoving sphere, consistently reproduces the Newtonian dynamical equation for a homogeneous self-gravitating medium. This result provides a convenient starting point for extending the description to more general fluids while keeping the underlying Newtonian structure transparent.

}

\

In order to include pressure 
$\bar{p}$ and describe a general fluid, one must account for the internal energy and its thermodynamic dependence on volume. This is accomplished by substituting the mass density 
$\bar{\sigma}$ with the homogeneous total energy density 
$\bar{\rho}$. The resulting modified Newtonian Lagrangian then takes the form~\cite{Harko}
{\begin{equation}
\bar{L}
= |{\bf r}|^2\left[\frac{1}{2}\left(\frac{da}{dt}\right)^2
+ \frac{4\pi G}{3}\, a^2 \bar{\rho}\right].
\end{equation}}

The Euler--Lagrange equation must now be supplemented by the first law of thermodynamics applied to a spherical volume of radius {$r=a|{\bf r}|$}:
\begin{equation}
d\!\left(\frac{4\pi}{3} r^3\bar{\rho}\right)
= -\bar{p}\, d\!\left(\frac{4\pi}{3} r^3\right)
\quad\Longleftrightarrow\quad
{d(a^3\bar{\rho}) = -\bar{p}\, d(a^3)}.
\end{equation}
Rewriting this expression yields
{\begin{equation}
a\, d(a^2\bar{\rho}) + a^2\bar{\rho}\, da = -3\bar{p}\, a^2\, da,
\end{equation}}
and therefore
{\begin{equation}
\frac{d}{da}(a^2\bar{\rho}) = -(\bar{\rho} + 3\bar{p})\, a.
\end{equation}}

Substitution into the 
{ Euler-Lagrange} equation 
{\begin{eqnarray}
    \frac{d^2 a}{dt^2}=\frac{4\pi G}{3}\frac{d}{da}(a^2\bar{\rho}),\end{eqnarray}}
leads to
\begin{equation}
{\frac{d^2 a}{dt^2}
= -\frac{4\pi G}{3}\,(\bar{\rho} + 3\bar{p})\, a}
\qquad\Longrightarrow\qquad
\frac{d^2 r}{dt^2}
= -\frac{G\,M_{\rm g}(r)}{r^2},
\end{equation}
where
\begin{equation}
M_{\rm g}(r)
= \frac{4\pi}{3}\, r^3\,(\bar{\rho} + 3\bar{p})
\end{equation}
defines the \textit{active gravitational mass}. This concept was introduced in Newtonian cosmology in~\cite{McCrea1951,Callan,
Harrison,McCrea1955} and later recognized within General Relativity in~\cite{Tolman}. The combination $\bar{\rho} + 3\bar{p}$ thus plays the role of the \textit{active gravitational mass density}.

When matter supports pressure, the Newtonian gravitational potential must be modified accordingly. In particular, the scalar potential $\Phi_1$ obeys the sourced wave equation
\begin{equation}
\Box \Phi_1 = -4\pi G\,(\rho + 3p),
\end{equation}
reflecting the fact that pressure contributes to gravitation through the active gravitational mass density $\rho(t,{\bf r})+3p(t,{\bf r})$.

\medskip

To place this result within a relativistic framework, consider the stress--energy tensor of a perfect fluid in Minkowski spacetime,
\begin{equation}
\mathfrak{T} = (\rho + p)\,{\bf u}^{\flat}\otimes{\bf u}^{\flat} - p\,\eta,
\end{equation}
where ${\bf u}^{\flat}$ denotes the one-form associated with the four-velocity ${\bf u}$.
The conservation law $\mathrm{div}(\mathfrak{T})=0$ yields the relativistic continuity and Euler equations,
\begin{equation}\label{relativisticconservationEuler}
\frac{d\rho}{ds} = -(\rho + p)\,\mathrm{div}({\bf u}),
\qquad
(\rho + p)\,\frac{d{\bf u}}{ds}
= \mathrm{grad}(p) - \frac{dp}{ds}\,{\bf u},
\end{equation}
where $(\rho+p)\,d{\bf u}/ds$ plays the role of the relativistic four-force {and 
$(\rho+p)\,{\bf u}$ the role of the four-momentum}. This identifies
\[
\rho_{\rm in}=\rho+p
\]
as the \textit{effective inertial mass density}, in the sense emphasized by Weinberg and Landau--Lifshitz~\cite{Weinberg,Landau}.

\medskip

These considerations naturally motivate, in the static weak-field regime, the introduction of a linearized invariant of the form
\begin{equation}
ds^2 = (1 + 2\Phi_1)\,dt^2 - (1 - 2\Phi_2)\,d{\bf r}\cdot d{\bf r},
\end{equation}
where the scalar potentials satisfy
\begin{equation}
\Box \Phi_1 = -4\pi G\,(\rho + 3p),
\qquad
\Box \Phi_2 = -4\pi G\,(\rho + b\,p),
\end{equation}
with $b$ a dimensionless parameter to be fixed by physical consistency.

\medskip

Applying the Lorentz transformation discussed {in the previous section, and retaining} terms linear in the velocity, the invariant becomes
\begin{equation}
ds^2 = (1 + 2\Phi_1)\,dt^2
- 8\,\bar{\bf N}\cdot d{\bf r}\,dt
- (1 - 2\Phi_2)\,d{\bf r}\cdot d{\bf r},
\end{equation}
where the vector potential satisfies
\begin{equation}\label{barN}
\Box \bar{\bf N}
= -4\pi G\!\left(\rho + \frac{1}{2}(3+b)p\right){\bf v}.
\end{equation}

{
Since the right-hand side of Eq.~(\ref{barN}) depends on 
$\left(\rho + \frac{1}{2}(3+b)p\right){\bf v}$, and since in the linear approximation the energy–momentum tensor is defined on the Minkowski background (see, e.g., Misner–Thorne–Wheeler \cite{MTW}), the relativistic momentum density takes the form $(\rho+p)\,{\bf v}$ for small velocities. 
Therefore, in order to assign a consistent physical interpretation to the term $\left(\rho + \frac{1}{2}(3+b)p\right){\bf v}$, we identify it with the three-momentum $(\rho+p){\bf v}$, which fixes
\[
b=-1.
\]

This shows that $\bar{\bf N}$ is sourced by the mass current density $(\rho+p){\bf v}$.

}

Consequently,
\begin{eqnarray}
\Box \Phi_2 = -4\pi G\,(\rho - p),
\qquad
\Box \bar{\bf N} = -4\pi G\,(\rho+p)\,{\bf v}.
\end{eqnarray}

\medskip

{

Including all velocity-dependent contributions, the invariant takes the fully linearized form
\begin{equation}\label{Invariant-full}
ds^2
= (1 - 2\Phi_2 + 4\hat{\Upsilon})\,dt^2
- 8\,\hat{\bf N}\cdot d{\bf r}\,dt
- (1 - 2\Phi_2)\,d{\bf r}\cdot d{\bf r}
+ 4\,\hat{\mathfrak{u}}(d{\bf r},d{\bf r}),
\end{equation}
with
\begin{equation}
\Box \hat{\Upsilon}
= -\frac{4\pi G(\rho+p)}{1-|{\bf v}|^2},\qquad
\Box \hat{\bf N}
= -\frac{4\pi G(\rho+p)}{1-|{\bf v}|^2}\,{\bf v},\qquad
\Box \hat{\mathfrak{u}}
= -\frac{4\pi G(\rho+p)}{1-|{\bf v}|^2}\,
{\bf v}^{\flat}\!\otimes{\bf v}^{\flat}.
\end{equation}

This invariant coincides exactly with the weak-field expression obtained in linearized General Relativity in harmonic gauge. Equivalently, it is governed by Einstein’s linear field equation (see Eq.~(96a)–(96b) of \cite{Einstein1955} or Eq.~(18.8b) of \cite{MTW}})
\begin{eqnarray}\label{linear}
\Box \mathfrak{h}
= -16\pi G \left(\mathfrak{T} - \frac{1}{2} \eta\, T \right), 
\qquad
ds^2=\eta(d{\bf x}, d{\bf x})+
\mathfrak{h}(d{\bf x}, d{\bf x}),
\end{eqnarray}
where the stress--energy tensor is defined in Minkowski spacetime and we use the notation ${\bf x}\equiv (t,{\bf r})$.

{
This correspondence can be verified as in \autoref{movingmasses} by explicitly computing the components of $T_{\mu\nu}-\frac{1}{2}\eta_{\mu\nu}T$ and checking that the coefficients of the invariant~\eqref{Invariant-full} satisfy the linearized equation~\eqref{linear}. For instance, for $\mu=\nu=0$ one finds
\begin{eqnarray}
T_{00}-\frac{1}{2}T
=\frac{\rho+p}{1-|{\bf v}|^2}-\frac{1}{2}(\rho-p),
\end{eqnarray}
which shows that the combination
$-2\Phi_2+4\hat{\Upsilon}$
indeed satisfies Eq.~\eqref{linear}.
}

\medskip

The potentials $\hat{\Upsilon}$, $\hat{\bf N}$, and $\hat{\mathfrak{u}}$ are sourced by the inertial mass density $\rho+p$. The interpretation of $\Phi_2$ is more subtle. Observing that
\[
\rho - p = 2(\rho+p) - (\rho+3p),
\]
we introduce auxiliary potentials $\hat{\Phi}_1$ and $\hat{\Upsilon}_1$ defined by
\[
\Box \hat{\Upsilon}_1
= -\frac{4\pi G(\rho+p)}{1-|{\bf v}|^2}|{\bf v}|^2,
\qquad
\Box \hat{\Phi}_1
= -4\pi G(\rho+p),
\]
which allow us to express
\begin{equation}
\Phi_2 = 2\hat{\Phi}_1 - \Phi_1.
\end{equation}

In terms of these quantities, the invariant may be rewritten as
\begin{equation}\label{Invariant-split}
ds^2
= (1 + 2(\Phi_1 + 2\hat{\Upsilon}_1))\,dt^2
- 8\,\hat{\bf N}\cdot d{\bf r}\,dt
- (1 + 2(\Phi_1 - 2\hat{\Phi}_1))\,d{\bf r}\cdot d{\bf r}
+ 4\,\hat{\mathfrak{u}}(d{\bf r},d{\bf r}).
\end{equation}

\medskip

Once the linearized invariant is known, the dynamics of the sources follow from the conservation law $\mathrm{div}(\mathfrak{T})=0$ in Minkowski spacetime, supplemented by an equation of state $p=p(\rho)$. This yields the relativistic continuity and Euler equations~\cite{Comer}:
\begin{equation}
\frac{d\rho}{ds}
= -(\rho+p)\,\mathrm{div}({\bf u}),
\qquad
(\rho+p)\,\frac{d{\bf u}}{ds}
= \mathrm{grad}(p) - \frac{dp}{ds}\,{\bf u}.
\end{equation}

In terms of the three-velocity ${\bf v}$, these equations become
\begin{eqnarray}
\partial_t\rho
+ \nabla\cdot\!\left((\rho+p)\frac{{\bf v}}{1-|{\bf v}|^2}\right)=0,
\nonumber\\
\partial_t\!\left((\rho+p)\frac{{\bf v}}{1-|{\bf v}|^2}\right)
+ \nabla\cdot\!\left((\rho+p)\frac{{\bf v}\otimes{\bf v}}{1-|{\bf v}|^2}\right)
+ \nabla p = 0.
\end{eqnarray}

Finally, taking the trace of Eq.~\eqref{linear} yields $\Box h = 16\pi G T$, allowing the field equation to be rewritten as
\begin{equation}
\Box\!\left(\mathfrak{h}-\frac{1}{2}\eta h\right)
= -16\pi G\,\mathfrak{T}.
\end{equation}
Imposing $\mathrm{div}(\mathfrak{T})=0$ then implies
\begin{equation}
\mathrm{div}\!\left(\mathfrak{h}-\frac{1}{2}\eta h\right)=0,
\end{equation}
which is precisely the linearized harmonic gauge condition. Conversely, in General Relativity formulated in harmonic gauge, the linearized equations automatically enforce energy--momentum conservation. In the weak-field regime, the harmonic gauge thus plays a role analogous to that of the Bianchi identities in the full theory~\cite{Bianchi}.

\medskip

In summary, starting from Galilean free fall, the operational notion of proper time introduced by Minkowski, and the extension of Fermat’s principle to massive bodies, we arrive—without postulating spacetime geometry or invoking {\color{red} the second} Newton’s law—at an invariant structure that coincides exactly with linearized General Relativity in harmonic gauge. As emphasized by Weinberg~\cite{Weinberg}:
\begin{quote}
\textit{All the classical tests of General Relativity—the precession of planetary orbits, the deflection of light, and the gravitational redshift—can be derived from the linearized field equations.}
\end{quote}

In this regime, gravity manifests itself neither as a Newtonian force nor as an abstract non-linear geometry, but as a physical modulation of proper time. The resulting description is conceptually direct, dynamically grounded, and fully sufficient to account for all classical gravitational phenomena.

\section{The general field equations in harmonic gauge}\label{Ricci}

We begin with the most general expression for the square of the differential of proper time,
\begin{equation}\label{propertime}
ds^2 = \frak{g}(d{\bf x},d{\bf x})=g_{\mu\nu}\,dx^{\mu}dx^{\nu},
\end{equation}
where we use the notation ${\bf x}\equiv(x^0, x^1, x^2, x^3)$.

\medskip

We now introduce the harmonic gauge (also known as the de Donder or Lorenz gauge).
To this end, we consider the natural generalization of the D’Alembert operator associated with the Minkowski invariant
\[
ds^2 =
\eta(d{\bf x},d{\bf x}),
\]
namely the Laplace--Beltrami operator associated with the general invariant (\ref{propertime}):
\begin{equation}
\Box_{\mathfrak{g}}\phi \equiv \operatorname{div}(\operatorname{grad}\phi)
= \frac{1}{\sqrt{-g}}\,
\partial_{\mu}\!\left(\sqrt{-g}\,g^{\mu\nu}\partial_{\nu}\phi \right),
\end{equation}
where $g$ denotes the determinant of the matrix $g_{\mu\nu}$.

\medskip

Harmonic coordinates are then defined as those satisfying
\begin{equation}
\Box_{\mathfrak{g}} x^{\mu}=0, \qquad \mu=0,1,2,3.
\end{equation}

In terms of the Christoffel symbols, the harmonic coordinate condition is equivalent to
\begin{equation}
g^{\mu\nu}\Gamma^{\alpha}_{\mu\nu}=0
\quad\Longleftrightarrow\quad
g^{\mu\nu}\!\left(\partial_{\mu}g_{\alpha\nu}
-\frac{1}{2}\partial_{\alpha}g_{\mu\nu}\right)=0
\quad\Longleftrightarrow\quad
\partial_{\mu}(\sqrt{-g}\,g^{\mu\alpha})=0,
\quad \alpha=0,1,2,3.
\end{equation}

In harmonic coordinates, the nonlinear generalization of the linear equation \eqref{linear} takes the form
\begin{equation}\label{generalization}
\Box_{\mathfrak{g}}\mathfrak{g} + \mathfrak{U}
=
-16\pi G \left( \mathfrak{T} - \frac{1}{2} \mathfrak{g} T \right),
\end{equation}
where $\Box_{\mathfrak{g}} = g^{\mu\nu}\partial_\mu \partial_\nu$, 
$\mathfrak{T}$ is the full matter energy--momentum tensor 
\begin{equation}
\frak{T} = (\rho + p) \, {\bf u}^\flat \otimes {\bf u}^\flat - p \, \frak{g}, 
\quad \text{with} \quad {\bf u} = \frac{d{\bf x}}{ds}, 
\quad ds = \sqrt{\frak{g}(d{\bf x}, d{\bf x})},
\end{equation}
and $\mathfrak{U}$ is a non--covariant term depending on $\mathfrak{g}$ and its first derivatives. 

This non-covariant term is unavoidable, because $\Box_{\mathfrak{g}}\mathfrak{g}$ by itself is not a tensor; only the combination $\Box_{\mathfrak{g}}\mathfrak{g} + \mathfrak{U}$ defines a genuine tensor. By construction, $\mathfrak{U}$ vanishes in the linear approximation.

Remarkably, in harmonic coordinates the Ricci tensor takes the form \cite{HE2026}
{\begin{eqnarray}
    R_{\mu\nu}&=&-\frac{1}{2}\Box_{\frak{g}}g_{\mu\nu}
-\frac{1}{2}(\partial_{\mu}g^{\alpha\beta}\partial_{\alpha}g_{\nu\beta}
+\partial_{\nu}g^{\alpha\beta}\partial_{\alpha}g_{\mu\beta})\nonumber\\ &&
-\frac{1}{4}g^{\beta\gamma}g^{\alpha\epsilon}\partial_{\mu}g_{\gamma\alpha}\partial_{\nu}g_{\beta\epsilon}
 -\frac{1}{2}g^{\beta \gamma}
    g^{\alpha \epsilon}\partial_{\alpha}g_{\mu\gamma}(\partial_{\beta}g_{\nu\epsilon}-\partial_{\epsilon}g_{\nu\beta}), \end{eqnarray}}
which we will denote as
\begin{equation}\label{Riccitensor}
\mathfrak{Ric}
=
-\frac{1}{2}\,\Box_{\mathfrak{g}} \mathfrak{g}
+\overline{\mathfrak{Ric}},
\end{equation}
where 
$\overline{\mathfrak{Ric}}$ denotes the non--covariant expression built from $\mathfrak{g}$ and its first derivatives.
This term collects all contributions quadratic in $\partial\mathfrak{g}$ and encodes the intrinsic nonlinearity of the gravitational field.

\vspace{0.2cm}

When $ds^2$ is a perturbation of Minkowski invariant, that is, 
$\mathfrak{g}=\eta+\mathfrak{h}$,
the linear approximation yields
\begin{equation}
\mathfrak{Ric} = -\frac{1}{2}\,\Box\,\mathfrak{h},
\end{equation}
which coincides exactly with the linearized field equations obtained previously.

{

It follows that the Ricci tensor provides the natural covariant nonlinear completion of the linearized field equation \eqref{linear}. More precisely, under the standard Lovelock assumptions—namely locality, dependence on the metric and its first two derivatives, second–order field equations, and compatibility with the covariant conservation of the energy–momentum tensor in four spacetime dimensions—the only symmetric rank–two tensor satisfying these requirements is the Ricci tensor (possibly supplemented by a cosmological term). Consequently, the field equations necessarily take the form
\begin{equation}\label{X}
\mathfrak{Ric}=
8\pi G \left(\mathfrak{T}-\frac{1}{2}\mathfrak{g}T\right),
\end{equation}
which coincide with the equations presented by Einstein in November 1915.

In this formulation, the Ricci tensor is not introduced as a prior geometric postulate; rather, it emerges as the unique covariant completion of the wave operator governing the propagation of the gravitational field once energy–momentum conservation and second–order dynamical consistency are imposed.

}

\

{

\begin{remark}
Alternatively, taking the trace of (\ref{generalization}) allows one to eliminate explicitly the scalar source $T$.
One obtains
\begin{eqnarray}
\Box_{\mathfrak{g}}\mathfrak{g}
+\mathfrak{U}
-\frac{1}{2}\,\mathfrak{g}\,
\big(\operatorname{Tr}(\Box_{\mathfrak{g}}\mathfrak{g})+U\big)
=
-16\pi G\,\mathfrak{T},
\end{eqnarray}
where $U\equiv \operatorname{Tr}(\mathfrak{U})$.
In this form, the right–hand side contains only the matter energy–momentum tensor, while all metric-dependent contributions are encoded on the left.

Taking the covariant divergence and using that we work in harmonic coordinates, all third–order derivatives of $\mathfrak{g}$ cancel identically. Imposing compatibility with energy–momentum conservation,
\begin{eqnarray}
\operatorname{div}(\mathfrak{T})=0,
\end{eqnarray}
one obtains the differential consistency condition
\begin{eqnarray}
\operatorname{div}\!\left(
\Box_{\mathfrak{g}}\mathfrak{g}
+\mathfrak{U}
-\frac{1}{2}\mathfrak{g}
\big(\operatorname{Tr}(\Box_{\mathfrak{g}}\mathfrak{g})+U\big)
\right)=0.
\end{eqnarray}


In four spacetime dimensions, Lovelock's theorem ensures that the only 
symmetric rank-two tensor built locally from $\mathfrak{g}$ and its first 
two derivatives, yielding second-order field equations and satisfying 
$\operatorname{div}(\mathfrak{T})=0$, is the Einstein tensor (up to a 
cosmological term). Therefore, the consistency condition above uniquely 
determines $\mathfrak{U}$, up to the cosmological  term $\lambda \frak{g}$, without the need for an explicit computation:
\begin{eqnarray}
\mathfrak{U}=-2\,\overline{\mathfrak{Ric}}.
\end{eqnarray}
Thus, the Ricci tensor emerges as the necessary nonlinear completion of 
the weak-field wave equation once energy–momentum conservation is imposed.
\end{remark}

}

\

Using \eqref{Riccitensor}, we may rewrite the Ricci tensor as
\begin{equation}
\mathfrak{Ric}
=
-\frac{1}{2}\Box_{\mathfrak{g}}\mathfrak{g}
+\left(\overline{\mathfrak{Ric}}-\frac{1}{2}\mathfrak{g}\,\bar{R}\right)
+\frac{1}{2}\mathfrak{g}\,\bar{R},
\end{equation}
where $\bar{R}$ is the trace of $\overline{\mathfrak{Ric}}$.

\

This motivates the introduction of the following non-covariant object,
\begin{equation}
\bar{\mathfrak{t}}
\equiv
-\frac{1}{8\pi G}
\left(\overline{\mathfrak{Ric}}-\frac{1}{2}\mathfrak{g}\,\bar{R}\right),
\end{equation}
which can be interpreted as the effective energy--momentum {\it pseudo--tensor} of the gravitational field.

With this definition, the Ricci tensor can be decomposed as follows:
\begin{eqnarray}
\frak{Ric}
=-\frac{1}{2}\Box_{\frak{g}}\frak{g}
-8\pi G \bar{\frak{t}}
+4\pi G\frak{g} \bar{t},
\end{eqnarray}
where now $\bar{t}$ denotes the trace of  $\bar{\frak{t}}$.

Substituting this expression into Einstein’s equations (\ref{X}), in harmonic coordinates, they take the form
\begin{eqnarray}
&&\Box_{\frak{g}}\frak{g}
= -16\pi G\left(
(\frak{T}+\bar{\frak{t}})
-\frac{1}{2}\frak{g} (T+\bar{t})
\right)
\Longleftrightarrow\nonumber\\&&
g^{\alpha\beta}\partial^2_{\alpha\beta}{g}_{\mu\nu}
= -16\pi G\left(
({T}_{\mu\nu}+\bar{{t}}_{\mu\nu})
-\frac{1}{2}{g}_{\mu\nu} (T+\bar{t})
\right).
\end{eqnarray}

This expression makes explicit that Einstein’s equations can be interpreted as a nonlinear wave equation where gravity does not merely respond to matter,
but also gravitates upon itself.
The self--interaction encoded in $\bar{\mathfrak{t}}$ is precisely the origin of the nonlinearity of General Relativity.

\

\begin{remark}
The expression of Einstein’s equations in harmonic coordinates allows for a clear dynamical interpretation of their physical content, going beyond the usual purely geometric formulation.
This structure is conceptually analogous to the dynamical formulation of gravitation developed by Feynman, Gupta, Thirring, and Deser \cite{Feynman,Gupta,Thirring,Deser}, in which General Relativity emerges as the consistent nonlinear completion of a field theory with massless spin--2 particles that couple universally to energy--momentum, including that of the field itself. In this sense, Einstein’s equations can be understood as the covariant and closed form that encapsulates all the necessary self-interactions of the gravitational field required to guarantee energy--momentum conservation and the dynamical consistency of the theory.

Thus, spacetime geometry does not appear as a fundamental starting point, but rather as the effective outcome of a dynamical field theory in which gravity gravitates upon itself.
\end{remark}

\

\begin{remark}
It is instructive to recall how Einstein arrived at his field equations \cite{Einstein1915,Einstein1916}, as this sheds light on the physical reasoning that ultimately guided their formulation.

{
In 1915 Einstein returned to the Ricci tensor after realizing that the equations of the {\it Entwurf} theory
\cite{Grossmann}
were unable to correctly account for the observed precession of Mercury’s perihelion. This observational failure forced him to reconsider the very foundations of his theory. Einstein then assumed that, in the absence of matter, the gravitational field equation must be
$$
\frak{Ric}=0.
$$
This choice proved decisive: under the assumption of spherical symmetry, and retaining terms up to second order, the equation reproduces exactly the observed perihelion precession of Mercury \cite{Einstein_Berlin}. This success marked the turning point that definitively convinced Einstein that he had found the correct path.

The next step consisted in determining the field equation in regions where matter is present. To address this problem, Einstein initially worked within what is now called {\it unimodular gravity}, that is, in coordinate systems satisfying
$$
g=-1,
$$
a condition that considerably simplifies the expression of the Ricci tensor. In these coordinates, and in the absence of matter, the equation
$$
\frak{Ric}=0,
$$
in coordinates, is equivalent to:
\begin{eqnarray}
\partial_{\alpha}\Gamma_{\mu\nu}^{\alpha}
=
\Gamma^{\alpha}_{\mu\beta}
\Gamma^{\beta}_{\nu\alpha},
\qquad \mbox{with} \quad g=-1.
\end{eqnarray}

From this expression, Einstein identified a conservation law of the form
$$
\partial_{\mu} t^{\mu}_{\nu}=0,
$$
where the pseudo--tensor $t^{\mu}_{\nu}$ is defined by
\begin{eqnarray}
8\pi G\, t^{\mu}_{\nu}
\equiv
\frac{1}{2}\delta_{\nu}^{\mu} g^{\alpha\sigma}
\Gamma^{\lambda}_{\alpha\beta}\Gamma^{\beta}_{\sigma\lambda}
-
g^{\alpha\sigma}\Gamma^{\mu}_{\alpha\beta}\Gamma^{\beta}_{\sigma\nu}.
\end{eqnarray}
By analogy with electrodynamics, Einstein interpreted this object as the energy--momentum pseudo--tensor of the gravitational field.

With this definition, the field equation in vacuum can be written as
\begin{eqnarray}
\partial_{\alpha}(g^{\nu\beta}\Gamma_{\mu\beta}^{\alpha})
=
8\pi G\left(
t^{\nu}_{\mu}
-
\frac{1}{2} \delta^{\nu}_{\mu} t
\right),
\qquad \mbox{with} \quad g=-1,
\end{eqnarray}
where $t\equiv t^{\alpha}_{\alpha}$ denotes the trace of the gravitational energy--momentum pseudo--tensor.

This expression is the key step for generalizing the field equation to regions containing matter. The reasoning is conceptually simple yet profound: the source of the gravitational field cannot be merely the energy of the gravitational field itself, but rather the total energy of the system, that is, the sum of the energy of matter and that of the gravitational field. In this way, Einstein obtained the complete field equation:
\begin{eqnarray}\label{Einstein_field}
\partial_{\alpha}(g^{\nu\beta}\Gamma_{\mu\beta}^{\alpha})
=
8\pi G\left(
(t^{\nu}_{\mu}+T^{\nu}_{\mu})
-
\frac{1}{2} \delta^{\nu}_{\mu}(t+T)
\right),
\qquad \mbox{with} \quad g=-1,
\end{eqnarray}
which automatically incorporates the conservation law of total energy--momentum:
$$
\partial_{\mu}(t^{\mu}_{\nu}+T^{\mu}_{\nu})=0,
$$
equivalent to the covariant condition
$$\operatorname{div}(\frak{T})=0.
$$

Finally, by regrouping terms so that only those associated with matter remain on the right-hand side, equation~(\ref{Einstein_field}) takes its definitive form:
\begin{eqnarray}
&&\partial_{\alpha}\Gamma_{\mu\nu}^{\alpha}
-
\Gamma^{\alpha}_{\mu\beta}\Gamma^{\beta}_{\nu\alpha}
=
8\pi G\left(
T_{\mu\nu}
-
\frac{1}{2} g_{\mu\nu} T
\right) \quad \mbox{with} \quad g=-1
\quad \Longleftrightarrow \nonumber\\ &&
\frak{Ric}
=
8\pi G\left(
\frak{T}
-
\frac{1}{2}\frak{g} T
\right),
\end{eqnarray}
which are nothing other than Einstein’s field equations in their fully covariant form.
}

\end{remark}

Ultimately, the results of this work invite a reconsideration of a deeply ingrained habit of thought: the tendency to regard gravity as a geometric structure superimposed upon an otherwise inert spacetime. The analysis presented here suggests a different reading. Gravity need not be postulated as an external geometric agency; it can emerge as the natural expression of a small set of physical principles whose simplicity is, in retrospect, striking. Galilean free fall, together with the extension of Fermat’s Principle to massive bodies—understood as the statement that proper time plays the role of a universal optical index—constitutes the conceptual core of this construction. In the absence of gravitation, physical trajectories extremize the Minkowski interval $ds$; when gravity is present, this invariant is locally modified, and it is the ensuing redistribution of proper time that governs motion.

From this strictly dynamical standpoint, Einstein’s field equations are not introduced as axioms nor as geometric postulates, but arise as the only coherent framework capable of accommodating gravitation within Special Relativity. Geometry, in this view, is not the ontological substrate of the theory, but its language: a codification of how physical clocks and rods respond to gravitational interaction. Within such a perspective, the harmonic gauge acquires a status that is no longer merely technical. It becomes the privileged arena in which the theory displays its physical content most transparently, revealing gravity as a modulation of the internal rhythm of time unfolding on a Minkowskian stage.

It is precisely in these coordinates that the weak-field regime appears in its clearest form, as the relativistic generalization of Newtonian gravity, without auxiliary hypotheses or interpretative artifices. In this sense, General Relativity returns to its conceptual origins: not as a theory about the curvature of an abstract manifold, but as a dynamical theory of time, inertia, and free fall. Dynamics, geometry, and the generalized Fermat principle for massive bodies thus converge into a unified and self-consistent interpretation, in which the mathematical structure of the theory reflects, rather than conceals, its physical and epistemological foundations.

\section{Linearized Cosmology in Harmonic Coordinates}

To describe an expanding or contracting universe it is convenient to introduce two types of coordinates: the physical (or proper) coordinates
\[
\mathbf{r}_{\rm ph}=(x_{\rm ph},y_{\rm ph},z_{\rm ph}),
\]
and the comoving coordinates
\[
\mathbf{r}=(x,y,z),
\]
related by
\begin{equation}
\mathbf{r}_{\rm ph}(t_{\rm c}) = a(t_{\rm c})\,\mathbf{r}(t_{\rm c}),
\end{equation}
where $t_{\rm c}$ is the \emph{cosmic time} and $a(t_{\rm c})$ is the \emph{scale factor}.

The physical velocity of a particle is defined as the time derivative of its physical position,
\begin{equation}
\mathbf{v}_{\rm ph}(t_{\rm c}) \equiv \frac{d\mathbf{r}_{\rm ph}}{dt_{\rm c}}
= \frac{d}{dt_{\rm c}}\!\left[a(t_{\rm c})\,\mathbf{r}(t_{\rm c})\right]
=  a'(t_{\rm c})\,\mathbf{r}(t_{\rm c})
+ a(t_{\rm c})\,{\mathbf{r}}'(t_{\rm c}),
\end{equation}
where a prime denotes differentiation with respect to $t_{\rm c}$.

The first term represents the velocity associated with the cosmological expansion,
\begin{equation}
\mathbf{v}_{\rm exp}(t_{\rm c})
\equiv  a'(t_{\rm c})\,\mathbf{r}(t_{\rm c})
= H(t_{\rm c})\,\mathbf{r}_{\rm ph}(t_{\rm c}),
\end{equation}
where
\[
H(t_{\rm c}) \equiv \frac{ a'(t_{\rm c})}{a(t_{\rm c})}
\]
is the Hubble rate.

The second term corresponds to the velocity with respect to comoving observers,
\begin{equation}
\mathbf{v}_{\rm co}(t_{\rm c})
\equiv a(t_{\rm c})\,{\mathbf{r}}'(t_{\rm c}).
\end{equation}

For the propagation of light one must impose that the comoving velocity has unit modulus,
\begin{equation}
|\mathbf{v}_{\rm co}| = 1,
\end{equation}
which implies
\begin{equation}
|{\mathbf{r}}'|^2 = \frac{1}{a^2(t_{\rm c})}.
\end{equation}
Explicitly,
\begin{equation}
\left(\frac{dx}{dt_{\rm c}}\right)^2
+\left(\frac{dy}{dt_{\rm c}}\right)^2
+\left(\frac{dz}{dt_{\rm c}}\right)^2
= \frac{1}{a^2(t_{\rm c})}.
\end{equation}
Multiplying by $dt_{\rm c}^2$, this condition can be written as
\begin{equation}
dt_{\rm c}^2 - a^2(t_{\rm c})\,d\mathbf{r}\cdot d\mathbf{r} = 0.
\end{equation}

By comparison with the propagation of light in Minkowski spacetime,
\[
dt^2 - d\mathbf{r}\cdot d\mathbf{r} = 0,
\]
one infers that the spacetime invariant in an expanding or contracting universe must be
\begin{equation}
ds^2 = dt_{\rm c}^2 - a^2(t_{\rm c})\,d\mathbf{r}\cdot d\mathbf{r},
\end{equation}
which coincides with the line element of the spatially flat
Friedmann--Lemaître--Robertson--Walker (FLRW) spacetime.

In harmonic coordinates $\mathbf{x}=(x^0=t,x^1,x^2,x^3)$, the invariant can be written as
\begin{equation}
ds^2 = \bar{\eta}_{\mu\nu}\,dx^\mu dx^\nu,
\end{equation}
with
\begin{equation}
\bar{\eta}_{\mu\nu}
= \mathrm{diag}\!\left(a^6(t),-a^2(t),-a^2(t),-a^2(t)\right).
\end{equation}


\

At the background level, in harmonic coordinates, the Ricci tensor reads
{\begin{equation}
R_{\mu\nu}^{(0)}
=-\frac{1}{2}\bar{\eta}^{00}\ddot{\bar{\eta}}_{\mu\nu}
-\delta_{\mu}^0\delta_{\nu}^0
\left(
\dot{\bar{\eta}}^{00}\dot{\bar{\eta}}_{00}
+\frac{1}{4}(\bar{\eta}^{\alpha\alpha})^2(\dot{\bar{\eta}}_{\alpha\alpha})^2
+\frac{1}{2}(\bar{\eta}^{00}\dot{\bar{\eta}}_{00})^2
\right)
+\frac{1}{2}\bar{\eta}^{\mu\nu}\bar{\eta}^{00}
(\dot{\bar{\eta}}_{\mu\nu})^2,
\end{equation}}
where the overdot denotes differentation with respect to the harmonic time $t$.

Therefore, one obtains
\begin{equation}
R_{kk}=\frac{1}{a^4}\left(\frac{\ddot a}{a}-H^2_{\rm har}\right),
\qquad
R_{00}=-3\left(\frac{\ddot a}{a}-3H^2_{\rm har}\right),
\qquad
R_{\mu\nu}=0\quad\text{for}\quad \mu\neq\nu ,
\end{equation}
where $H_{\rm har}=\dot{a}/a$ is the Hubble rate in harmonic coordinates.

On the other hand, the stress-energy tensor takes the form
\begin{equation}
T^{\mu}{}_{\mu}=\bar{\rho}-3\bar{p},
\qquad
T_{kk}=\bar{p} a^2,
\qquad
T_{00}=\bar{\rho} a^6,
\qquad
T_{\mu\nu}=0\quad\text{for}\quad \mu\neq\nu ,
\end{equation}
where $\bar{\rho}$ and $\bar{p}$, once again,  denote the homogeneous energy density and pressure,
respectively.

For $k=1,2,3$, Einstein’s equations yield
\begin{equation}
 {\dot{H}_{\rm har} =}\frac{\ddot a}{a}-H^2_{\rm har}
=
4\pi G\,a^6(\bar{\rho}-\bar{p}),
\end{equation}
whereas for $\mu=0$ one obtains the second Friedmann equation in harmonic
coordinates,
\begin{equation}
{-3\dot{H}_{\rm har}+6H_{\rm har}^2=}-3\frac{\ddot a}{a}+9H^2_{\rm har}
=
4\pi G\,a^6(\bar{\rho}+3\bar{p}).
\end{equation}

Combining both equations leads to the first Friedmann equation in harmonic
coordinates,
\begin{eqnarray}
H^2_{\rm har}=\frac{8\pi G}{3}a^{6}\bar{\rho} .
\end{eqnarray}

\begin{remark}

Expressed in terms of the cosmic time $t_{\rm c}$, the Friedmann equations take their standard form,
    \begin{eqnarray}
    H^2=\frac{8\pi G}{3}\bar{\rho},\qquad \frac{a''}{a}=-\frac{4\pi G}{3}(\bar{\rho}+3\bar{p}).
\end{eqnarray}

Using the relation between cosmic time and harmonic time,
\begin{equation}
dt_{\rm c}=a^3(t)\,dt,
\end{equation}
these equations can be straightforwardly rewritten in harmonic coordinates, yielding the Friedmann equations in harmonic gauge.

\end{remark}

\subsection{Cosmic perturbations}

In the presence of cosmological perturbations, we write
\begin{equation}
g_{\mu\nu}=\bar{\eta}_{\mu\nu}+h_{\mu\nu}+\mathcal{O}(h^2),
\end{equation}
and for the inverse metric,
\begin{equation}
g^{\mu\nu}=\bar{\eta}^{\mu\nu}-h^{\mu\nu}+\mathcal{O}(h^2),
\end{equation}
with $\bar{\eta}_{\mu\alpha}\bar{\eta}^{\alpha\nu}=\delta_{\mu}^{\nu}$.
Using $g_{\mu\alpha}g^{\alpha\nu}=\delta_{\mu}^{\nu}$, one finds
\begin{equation}
h^{\mu\nu}
=
\bar{\eta}^{\mu\alpha}\bar{\eta}^{\nu\beta}h_{\alpha\beta}.
\end{equation}

The linearized harmonic gauge condition then becomes
\begin{equation}
\bar{\eta}^{\alpha\alpha}
\left(
\partial_{\alpha}h_{\mu\alpha}
-\frac{1}{2}\partial_{\mu}h_{\alpha\alpha}
\right)
+h^{\mu 0}\dot{\bar{\eta}}_{\mu\mu}
-\frac{1}{2}\delta_{\mu}^0 h^{\alpha\alpha}\dot{\bar{\eta}}_{\alpha\alpha}
=0,
\qquad
\mu=0,1,2,3 .
\end{equation}

At first order in perturbations, the Ricci tensor can be written as
{\begin{eqnarray}
    R_{\mu\nu}^{(1)}&=&-\frac{1}{2}\Box_{\rm har}h_{\mu\nu}
-\frac{1}{2}\delta_{\mu}^{\alpha}\delta_{\nu}^{\alpha}h^{00}\ddot{\bar{\eta}}_{\alpha\alpha}
    -\mbox{Sym}\left[\delta_{\mu}^0\dot{\bar{\eta}}^{\alpha\alpha}\partial_{\alpha}h_{\nu\alpha}+\partial_{\mu}h^{0\nu}\dot{\bar{\eta}}_{\nu\nu}
    \right]-\frac{1}{2}\delta_{\mu}^0\delta_{\nu}^0h^{\alpha\alpha}\bar{\eta}^{\alpha\alpha}(\dot{\bar{\eta}}_{\alpha\alpha})^2\nonumber\\
   &&-\frac{1}{2}\mbox{Sym}\left[ 
\delta_{\nu}^0(\bar{\eta}^{\alpha\alpha})^2\dot{\bar{\eta}}_{\alpha\alpha}\partial_{\mu}h_{\alpha\alpha}
    \right]
-\bar{\eta}^{00}\dot{\bar{\eta}}_{00}
\mbox{Sym}[\dot{\bar{\eta}}_{\mu\mu}\delta_{\nu}^0 h^{0\mu}
]
\nonumber\\
&&+
\frac{1}{2}\dot{\bar{\eta}}_{\mu\mu}\dot{\bar{\eta}}_{\nu\nu}
(\bar{\eta}^{00}h^{\mu\nu}+\bar{\eta}^{\mu\nu}h^{00})
-
\bar{\eta}^{00}\mbox{Sym}[\bar{\eta}^{\nu\nu}\dot{\bar{\eta}}_{\nu\nu}(\partial_{\nu}h_{\mu 0}-\dot{h}_{\mu\nu})]
\nonumber\\ &\equiv&
 -\frac{1}{2}\Box_{\rm har}h_{\mu\nu}+ F_{\mu\nu}^{(1)}(\frak{h}, \partial\frak{h}),
 \end{eqnarray}}
where
\[
\Box_{\rm har}
=
\bar{\eta}^{\alpha\beta}\partial_{\alpha}\partial_{\beta}
=
\frac{1}{a^6}\partial_{tt}^2-\frac{1}{a^2}\Delta,
\]
and $\mathrm{Sym}$ denotes symmetrization with respect to the indices $\mu$
and $\nu$.

Writing $\rho=\bar{\rho}+\delta{\rho}$ and $p=\bar{p}+\delta{p}$, and treating the spatial
velocities $\dot{x}^k$ ($k=1,2,3$) as perturbative quantities, one finds at
first order
\begin{eqnarray}
(g_{\mu\nu}T)^{(1)}
=
\bar{\eta}_{\mu\nu}(\delta{\rho}-3\delta{p})
+h_{\mu\nu}\bar{T},
\end{eqnarray}
where $\bar{T}=\bar{\rho}-3\bar{p}$ is the background trace of the stress-energy tensor.
Moreover,
{\begin{eqnarray}
T_{\mu\nu}^{(1)}
&=&
(\delta{\rho}+\delta{p})\bar{\eta}_{00}\delta_{\mu}^{ 0}\delta_{\nu}^{ 0}
+2(\bar{\rho}+\bar{p})
\mathrm{Sym}[\delta_{\mu}^{0}\dot{x}_{\nu}-\bar{\eta}_{00}\delta_{\mu}^{0}\delta_{\nu}^{0}]\nonumber\\&&
-(\bar{\rho}+\bar{p})
\delta_{\mu}^{ 0}\delta_{\nu}^{ 0} h_{00}
-
\bar{\eta}_{\mu\nu}\delta{p}
-h_{\mu\nu}\bar{p} .
\end{eqnarray}}


Inserting these expressions into the first-order perturbed Einstein equations,  
\begin{eqnarray}
-\frac{1}{2}\Box_{\rm har}h_{\mu\nu} + F_{\mu\nu}^{(1)}(\mathfrak{h}, \partial\mathfrak{h}) =
8\pi G \left( T_{\mu\nu}^{(1)} - \frac{1}{2}(g_{\mu\nu}T)^{(1)} \right),
\end{eqnarray}  
yields the linearized field equations in harmonic coordinates.  
These equations take the form of a linear wave equation with source \(T_{\mu\nu}^{(1)}\), while the term 
\(F_{\mu\nu}^{(1)}(\mathfrak{h}, \partial\mathfrak{h})\) can be interpreted as an effective damping or “friction” term arising from the background expansion, encoding additional contributions from the time dependence of the metric.

\

Finally, in the limit in which the cosmological expansion can be neglected,
namely $a=1$ and $\bar{\rho}=\bar{p}=0$, one recovers the standard linearized Einstein
equations around Minkowski spacetime, as expected.

\

{

\subsection{Scalar perturbations}

After a straightforward but lengthy calculation, one obtains:
{\small \begin{eqnarray}\left\{\begin{array}{ccc}
   \sum_{k=1}^3 R_{kk}^{(1)}&=&-\frac{1}{2}\Box_{\rm har}(\sum_{k=1}^3   h_{kk})+
    \frac{3}{a^{10}}\dot{H}_{\rm har}h_{00}+\frac{2}{a^6}H_{\rm har}^2\sum_{k=1}^3h_{kk}-\frac{4}{a^6}H_{\rm har}\sum_{k=1}^3    \left(\partial_kh_{k0}-\frac{1}{2}\dot{h}_{kk}
\right),\\&&\\
    R_{00}^{(1)}&=&-\frac{1}{2}\Box_{\rm har}h_{00}
 -\frac{3}{a^6}\frac{d(H_{\rm har}{h}_{00})}{dt}   
    +\frac{36}{a^6}H^2_{\rm har}h_{00}+\frac{2}{a^2}H^2_{\rm har}h_{kk}
    -\frac{2}{a^2}H_{\rm har}\sum_{k=1}^3    \left(\partial_k h_{0k}
-\frac{1}{2}\dot{h}_{kk}\right).\end{array}\right.
    \end{eqnarray}}

Taking into account that
the 
linear harmonic gauge for $\mu=0$, leads to:
\begin{eqnarray}\sum_{k=1}^3   (\partial_k h_{k0}-\frac{1}{2}\dot{h}_{kk})=\frac{1}{2a^4}\dot{h}_{00}+\frac{3}{a^4}H_{\rm har}h_{00}+ H_{\rm har}
\sum_{k=1}^3   h_{kk},
\end{eqnarray}
the previous expressions can be written as:
{\small\begin{eqnarray}\left\{\begin{array}{ccc}
\sum_{k=1}^3 R_{kk}^{(1)}&=&-\frac{1}{2}\Box_{\rm har}
(\sum_{k=1}^3 h_{kk})+
    \frac{3}{a^{10}}\dot{H}_{\rm har}h_{00}
    -\frac{2}{a^{10}}H_{\rm har}\dot{h}_{00}
    -\frac{12}{a^{10}}H^2_{\rm har}h_{00}
    -\frac{2}{a^6}H_{\rm har}^2\sum_{k=1}^3 h_{kk},
    \\&&\\
    R_{00}^{(1)}&=&-\frac{1}{2}\Box_{\rm har}h_{00}
 -\frac{3}{a^6}{\dot{H}_{\rm har}{h}_{00}}   
    +\frac{30}{a^6}H^2_{\rm har}h_{00}-\frac{4}{a^6}H_{\rm har}\dot{h}_{00}.\end{array}\right.
    \end{eqnarray}}

Next, 
we introduce the potentials $h_{00}\equiv 2a^6\Psi_1$ and 
$\sum_{k=1}^3 h_{kk}\equiv 6a^2\Psi_2$.
Then, we have:
\begin{eqnarray}\left\{\begin{array}{ccc}
\sum_{k=1}^3 R_{kk}^{(1)}&=&-3\Box_{\rm har}(a^2\Psi_2)+\frac{6}{a^4}\dot{H}_{\rm har}\Psi_1-\frac{48}{a^4}H^2_{\rm har}\Psi_1-\frac{4}{a^4}H_{\rm har}\dot{\Psi}_1-\frac{12}{a^4}H^2_{\rm har}\Psi_2,
\\&&\\
    R_{00}^{(1)}&=&-\Box_{\rm har}(a^6\Psi_1)-6\dot{H}_{\rm har}\Psi_1+12 H_{\rm bar}^2\Psi_1-8H_{\rm har}\dot{\Psi}_1.\end{array}\right.
\end{eqnarray}

For the 
stress-tensor, in the linear approximation, one has:
\begin{eqnarray}\left\{\begin{array}{ccc}
  \sum_{k=1}^3 \left( T_{kk}^{(1)}-\frac{1}{2}(g_{kk}T)^{(1)}\right)
    &=&\frac{3}{2}a^2(\delta\rho-\delta p)-3a^2(\bar{\rho}-\bar{p})\Psi_2,
    \\&&\\
    T_{00}^{(1)}-\frac{1}{2}(g_{00}T)^{(1)}    
    &=&\frac{1}{2}a^6(\delta \rho+3\delta p) -{3}a^6(\bar{\rho}-\bar{p})\Psi_1.
\end{array}\right.\end{eqnarray}

Therefore, the dynamical equations are: 
{\small\begin{eqnarray}\left\{\begin{array}{ccc}
-3a^4\Box_{\rm har}(a^2\Psi_2)+{6}\dot{H}_{\rm har}\Psi_1-{48}H^2_{\rm har}\Psi_1-{4}H_{\rm har}\dot{\Psi}_1-{12}H^2_{\rm har}\Psi_2
&=&12 \pi Ga^6(\delta\rho-\delta p)-24 \pi Ga^6(\bar{\rho}-\bar{p})\Psi_2, \\&&\\
    -\Box_{\rm har}(a^6\Psi_1)-6\dot{H}_{\rm har}\Psi_1+12 H_{\rm har}^2\Psi_1-8H_{\rm har}\dot{\Psi}_1&=&
    4\pi G a^6(\delta \rho+3\delta p) -{24 \pi G}a^6(\bar{\rho}-\bar{p})\Psi_1. \end{array}\right.   
\end{eqnarray}}

And using the background equation
\begin{eqnarray}
    \dot{H}_{\rm har}=4\pi G a^6(\bar{\rho}-\bar{p}),
\end{eqnarray}
these equations become:
{\small\begin{eqnarray}\left\{\begin{array}{ccc}
-3\Box_{\rm har}(a^6\Psi_1)
    +36 H_{\rm har}^2\Psi_1-24H_{\rm har}\dot{\Psi}_1 &=&
    12\pi G a^6(\delta \rho+3\delta p), 
    \\&&\\
-3a^4\Box_{\rm har}(a^2\Psi_2)+{6}\dot{H}_{\rm har}(\Psi_1+\Psi_2)-{12}H^2_{\rm har}(4\Psi_1+\Psi_2)-{4}H_{\rm har}\dot{\Psi}_1
&=&12 \pi Ga^6(\delta\rho-\delta p).
    \end{array}\right.
\end{eqnarray}}

We can thus identify the potentials $\Psi_1$ and $\Psi_2$ 
as the natural generalizations, in an expanding universe, of the static potentials
 $\Phi_1$ and $\Phi_2$
 introduced in Sec.~\ref{sec-C}. In the static limit 
 $H_{\rm har}=0$,  the two sets of potentials coincide,
 {because, when $H_{\rm har}=0$, these equations become
\begin{eqnarray}
\Box\Psi_1
    =
    -4\pi G (\delta \rho+3\delta p), 
    \qquad
\Box\Psi_2
=-4\pi G (\delta\rho-\delta p),
\end{eqnarray}
which coincide with the equations satisfied by the potentials $\Phi_1$ and $\Phi_2$.
}

}

\

A final remark is in order.
{\it 
In cosmology, perturbations are usually decomposed into scalar, vector, and tensor modes. 
The perturbations directly related to the gravitational potential are the scalar ones. 
In the so-called Newtonian gauge, the gauge-invariant metric perturbation takes the form
\begin{eqnarray}
    ds^2=(1+2\Phi_{\rm N})dt_{\rm c}^2-a^2(t_{\rm c})(1-2\Phi_{\rm N})d{\bf r}\cdot d{\bf r},
\end{eqnarray}
where $\Phi_{\rm N}$ denotes the Newtonian potential.

The corresponding perturbation equations are well known (see, for instance, Eqs.~(7.47)–(7.49) in~\cite{Mukhanov}) and, in a static universe, where $t_{\rm c}=t$, reduce to the simple expressions
\begin{eqnarray}
    \Delta\Phi_{\rm N}=4\pi G \delta{\rho}, 
    \qquad 
    \ddot{\Phi}_{\rm N}=4\pi G\delta{p}, 
    \qquad 
    \nabla\dot{\Phi}_{\rm N}={\bf 0}.
\end{eqnarray}

From these equations, however, the interpretation of $\Phi_{\rm N}$ as a genuine gravitational potential is at best indirect.
Although $\Phi_{\rm N}$ is a gauge-invariant quantity, it does not emerge as a dynamical potential governing the motion of test particles in a direct Newtonian sense.
Rather, it appears as a metric perturbation whose physical meaning becomes transparent only after fixing a specific gauge and restricting to particular regimes.

This stands in contrast with the clear dynamical interpretation of the potentials $\Phi_1$ and $\Phi_2$ obtained in Section~\ref{sec-C}, which arise naturally in harmonic coordinates and are directly related to the equations of motion.
}

{

\subsection{Vector and tensor perturbations}

We start with the linearized Ricci tensor components:
\begin{align}
    R_{k0}^{(1)} &= -\frac{1}{2}\Box_{\rm har} h_{k0}
    -\frac{1}{2}\Big(\partial_k h^{00}\dot{\bar{\eta}}_{00} + \dot{h}^{0k} \dot{\bar{\eta}}_{kk} \Big)
    - \frac{1}{2} \bar{\eta}^{00} \bar{\eta}^{kk} \dot{\bar{\eta}}_{kk} (\partial_k h_{00} - \dot{h}_{k0}) \nonumber\\
    &\quad - \frac{1}{2} \Big( \dot{\bar{\eta}}^{\alpha\alpha} \partial_\alpha h_{k\alpha} + \frac{1}{2} (\bar{\eta}^{\alpha\alpha})^2 \dot{\bar{\eta}}_{\alpha\alpha} \partial_k h_{\alpha\alpha} \Big).
\end{align}

Using the background quantities, this reduces to
\begin{align}
    R_{k0}^{(1)} &= -\frac{1}{2}\Box_{\rm har} h_{k0}
    -\frac{H_{\rm har}}{a^6}\Big(3\partial_k h_{00} - \dot{h}_{0k} + 8 H_{\rm har} h_{k0}\Big)
    - \frac{H_{\rm har}}{a^6} (\partial_k h_{00} - \dot{h}_{k0}) \nonumber\\
    &\quad + \frac{3 H_{\rm har}}{a^6} \Big(\dot{h}_{k0} - \frac{1}{2} \partial_k h_{00} \Big)
    + \frac{H_{\rm har}}{a^2} \sum_{i=1}^3\Big(\partial_i h_{ki} - \frac{1}{2} \partial_k h_{ii} \Big).
\end{align}

Next, applying the harmonic gauge condition:
\begin{align}
    \bar{\eta}^{\alpha\alpha} \Big( \partial_\alpha h_{k\alpha} - \frac{1}{2} \partial_k h_{\alpha\alpha} \Big) + h^{k0} \dot{\bar{\eta}}_{kk} = 0,
\end{align}
we can write
\begin{align}
    \frac{1}{a^2} \sum_{i=1}^3\Big( \partial_i h_{ki} - \frac{1}{2} \partial_k h_{ii} \Big)
    = \frac{1}{a^6} \Big( \dot{h}_{k0} - \frac{1}{2} \partial_k h_{00} + 2 H_{\rm har} h_{k0} \Big).
\end{align}

Combining the above, one obtains
\begin{align}
    R_{k0}^{(1)} = -\frac{1}{2}\Box_{\rm har} h_{k0} - \frac{6 H_{\rm har}}{a^6} \big(\partial_k h_{00} - \dot{h}_{k0} - H_{\rm har} h_{k0}\big).
\end{align}

On the other hand, for the stress-energy tensor:
\begin{align}
    T_{k0}^{(1)} - \frac{1}{2} (g_{k0} T)^{(1)} = -\frac{1}{2} (\bar{\rho} - \bar{p}) h_{k0} + (\bar{\rho} + \bar{p}) \dot{x}_k.
\end{align}

Introducing the notation $h_{k0} = 4 a^6 N_k$, the linearized Einstein equations give
\begin{align}
    -2 \Box_{\rm har} (a^6 N_k) - 12 H_{\rm har} \Big( \partial_k \Psi_1 - 14 H_{\rm har} N_k - 2 \dot{N}_k \Big)
    = -4 \pi G (\bar{\rho}-\bar{p}) h_{k0} + 8 \pi G (\bar{\rho} + \bar{p}) \dot{x}_k.
\end{align}

Using the background equation $\dot{H}_{\rm har} = 4 \pi G a^6 (\bar{\rho} - \bar{p})$, we finally write in vector notation:
\begin{align}
    -\Box_{\rm har} (a^6 \mathbf{N}_{\rm har}) - 6 H_{\rm har} (\nabla \Psi_1 - 14 H_{\rm har} \mathbf{N}_{\rm har} - 2 \dot{\mathbf{N}}_{\rm har})
    + 2 \dot{H}_{\rm har} \mathbf{N}_{\rm har} = 4 \pi G a^6(\bar{\rho} + \bar{p}) \mathbf{v}_{\rm har},
\end{align}
where $\mathbf{N}_{\rm har} = (N_1, N_2, N_3)$ and ${\bf v}_{\rm har}=\frac{1}{a^6}(\dot{x}_1, \dot{x}_2, \dot{x}_3)$.

\

For tensor perturbations with $i \neq k$, we have
\begin{eqnarray}\left\{\begin{array}{ccc}
    R_{ik}^{(1)} &=& -\frac{1}{2} \Box_{\rm har} h_{ik} + \frac{2 H_{\rm har}^2}{a^6} h_{ik} 
    - \frac{2 H_{\rm har}}{a^6} (\partial_i h_{k0} + \partial_k h_{i0} - \dot{h}_{ik}),
\\&&\\
    T_{ik}^{(1)} - \frac{1}{2} (g_{ik} T)^{(1)} &=& -\frac{1}{2} (\bar{\rho} - \bar{p}) h_{ik}.\end{array}\right.
\end{eqnarray}

Defining $h_{ik} = 4 a^6 u_{ik}$ and using again $\dot{H}_{\rm har} = 4 \pi G a^6 (\bar{\rho}-\bar{p})$, we obtain
\begin{align}
    -\Box_{\rm har} (a^6 u_{ik}) + 4 H_{\rm har}^2 u_{ik} - 4 H_{\rm har} (\partial_i N_k + \partial_k N_i - \dot{u}_{ik})
    + 2 \dot{H}_{\rm har} u_{ik} = 0.
\end{align}

Introducing the tensor $\frak{u}_{\rm har}$ with components $u_{ik}$ for $i \neq k$ and $u_{kk} = 0$, we can write
\begin{align}
    -\Box_{\rm har} (a^6 \frak{u}_{\rm har}) + 4 H_{\rm har}^2 \frak{u}_{\rm har} - 4 H_{\rm har} (2 \frak{N} - \dot{\frak{u}}_{\rm har})
    + 2 \dot{H}_{\rm har} \frak{u}_{\rm har} = 0,
\end{align}
where
\begin{align}
    \frak{N} \equiv \frac{1}{2} \left(\nabla \mathbf{N}_{\rm har} + (\nabla \mathbf{N}_{\rm har})^T\right)
              -  \mathrm{diag} \big( \nabla \mathbf{N}_{\rm har} \big),
\end{align}
is the \emph{pure shear tensor}, i.e., the off-diagonal symmetric part of the gradient of $\mathbf{N}_{\rm har}$.

\

Therefore, the perturbation equations can be summarized as follows:
{\small\begin{eqnarray}
    \left\{\begin{array}{ccc}
-3\Box_{\rm har}(a^6\Psi_1)
    +36 H_{\rm har}^2\Psi_1-24H_{\rm har}\dot{\Psi}_1 &=&
    12\pi G a^6(\delta \rho+3\delta p), 
\\ & &
    \\
-3a^4\Box_{\rm har}(a^2\Psi_2)+{6}\dot{H}_{\rm har}(\Psi_1+\Psi_2)-{12}H^2_{\rm har}(4\Psi_1+\Psi_2)-{4}H_{\rm har}\dot{\Psi}_1
&=&12 \pi Ga^6(\delta\rho-\delta p),\\ & &\\
-3\Box_{\rm har} (a^6 \mathbf{N}_{\rm har}) - 12 H_{\rm har} (\nabla \Psi_1 - 14 H_{\rm har} \mathbf{N}_{\rm har} - 2 \dot{\mathbf{N}}_{\rm har})
    + 6 \dot{H}_{\rm har} \mathbf{N}_{\rm har} &=& 12 \pi G a^6(\bar{\rho} + \bar{p}) \mathbf{v}_{\rm har},\\ & &\\
      -3\Box_{\rm har} (a^6 \frak{u}_{\rm har}) + 12 H_{\rm har}^2 \frak{u}_{\rm har} - 12 H_{\rm har} (2 \frak{N} - \dot{\frak{u}}_{\rm har})
    + 6 \dot{H}_{\rm har} \frak{u}_{\rm har}&=& 0.
    \end{array}\right.
\end{eqnarray}}

\

Finally, the linear harmonic gauge yields the following relations among the
gravitational potentials:
\begin{eqnarray}\left\{\begin{array}{ccc}
    4 a^4 \nabla \cdot {\bf N}_{\rm har}
    &=&
    12 H_{\rm har} (\Psi_1 + \Psi_2)
    + \dot{\Psi}_1 + 3 \dot{\Psi}_2 ,
\\ &&\\
    4 a^4 \nabla \cdot \frak{u}_{\rm har}
    &=&
    32 H_{\rm har} {\bf N}_{\rm har}
    + 4 \dot{\bf N}_{\rm har}
    + \nabla (3 \Psi_2 - \Psi_1) .
    \end{array}\right.
\end{eqnarray}
Here $\nabla \cdot \frak{u}_{\rm har}$ denotes the three--divergence of $\frak{u}_{\rm har}$, i.e.,
the vector with components
\[
(\nabla \cdot \frak{u}_{\rm har})_k \equiv \sum_{i=1}^3 \partial_i u_{ik}.
\]
The equations of motion for matter—the relativistic continuity and Euler
equations—are obtained from the linearized conservation law
$\mathrm{div}(\frak{T}) = 0$.

}


\section{Conclusions}

The construction of General Relativity emerged from a profound and nontrivial dialogue between physical intuition and mathematical structure. From Minkowski’s unification of space and time to Einstein’s sustained efforts between 1907 and 1912 to account for gravitation through the invariant $ds^2$, the guiding principles were consistently physical in nature: the Equivalence Principle, the conservation of energy and momentum, and the Principle of Relativity. Geometry entered decisively only in 1913, when Grossmann introduced Einstein to Riemannian geometry—a development that ultimately culminated in the field equations of 1915. Without these mathematical tools, the final formulation of General Relativity would undoubtedly not have been possible. Yet this success came with a price: a gradual shift of emphasis from physical reasoning to formal structure, a shift whose interpretative consequences still permeate the standard understanding of the theory.

The present work develops a complementary standpoint—one that Einstein himself might plausibly have pursued prior to 1913—in which the invariant $ds^2$ is not postulated geometrically, but determined dynamically by matter through Lorentz invariance and an extension of Fermat’s principle. In this framework, spacetime curvature is not assumed as a primitive notion. The central object is instead the invariant $ds$ itself and its integral along a worldline,
$s=\int ds$.

In Special Relativity, this integral coincides with the proper time experienced along a trajectory. In a gravitational context, proper time is modified in such a way that freely falling particles follow stationary paths of $s$. Gravitation thus manifests itself as a modulation of the local flow of proper time, simultaneously governing the motion of bodies and the rate at which clocks tick. Dynamics and temporal measurement are thereby unified at a fundamental level.

From this perspective, the geometrization of gravity acquires a different interpretation. The invariant $ds$ encodes how proper time is operationally defined, while curvature—or, more generally, geometric structures—serve as efficient mathematical tools for describing the relational evolution of matter and fields. What is physically essential is not the particular geometric representation adopted, but the internal consistency of the dynamics: the mutual coherence between particle motion, clock rates, and conservation laws. In this sense, Einstein’s own caution against reifying spacetime should be read as a warning against conflating mathematical representation with physical ontology.

As shown explicitly in the weak-field regime, this approach naturally reproduces the Newtonian limit. 
Once the weak-field limit is firmly established, the resulting equations are found to coincide precisely with Einstein’s equations linearized in harmonic gauge. This observation is crucial: it shows that, even without postulating curvature from the outset, the theory contains an intrinsic pathway toward its extension into the strong-field regime. The structure of the Ricci tensor in harmonic coordinates provides exactly the mathematical guidance required to extend the weak-field dynamics in a consistent and closed manner.

This extension is not arbitrary. Lovelock’s theorem demonstrates that, in four spacetime dimensions and under broad and physically natural assumptions—namely, second-order field equations and conservation of the energy--momentum tensor—the Ricci tensor yields the unique non-linear completion of the linearized theory. Spacetime curvature thus emerges not as a metaphysical hypothesis, but as the only mathematically consistent mechanism for encoding gravitational self-interaction beyond the weak-field approximation.

Seen in this light, the present perspective does not contradict Einstein’s original intuition that gravitation reflects a modification of inertial structure. On the contrary, it brings that intuition to completion: once dynamical consistency is required, such a modification can only be captured by the geometric framework Einstein ultimately adopted in 1915. Curvature appears, therefore, as the inevitable mathematical expression of gravity when the weak-field description is pushed to its logical conclusion.

{

It is also worth emphasizing that alternative but dynamically equivalent formulations of gravitation—most notably teleparallel gravity and other tetrad- or frame-based approaches—likewise attribute a central role to the interplay between inertial and gravitational effects through the choice of reference frame. Because the present construction reproduces exactly the weak-field equations of General Relativity in harmonic gauge, its physical predictions in this regime are necessarily consistent with those obtained in teleparallel gravity, which is known to be equivalent to General Relativity at the level of the field equations. However, the purpose of the present work has been to remain within a purely metric and dynamical framework based on proper time and Lorentz invariance. A reformulation in terms of proper frames or tetrads—particularly in order to further clarify the separation between inertial and gravitational contributions—would constitute a natural and interesting direction for future research.

}

In summary, the weak-field limit of General Relativity admits two complementary readings: a geometric one, centered on curved spacetime, and a dynamical one, focused on the interaction between matter, inertia, and proper time through the invariant $ds$. While the geometric picture remains powerful and elegant, it does not exhaust the physical content of the theory. By recovering General Relativity linearized in harmonic gauge, the approach developed here
{restores the primacy of physical principles and offers a transparent interpretation of gravitation in the weak-field regime. Geometry is not a postulate but a consequence: it enters the theory only when dynamical consistency demands a non-linear extension of the weak-field equations, at which point the Ricci tensor emerges as the unique available structure. In this sense, General Relativity is a geometric theory — but geometry is its outcome, not its starting point.
}

{
Ultimately, this dynamical coherence between matter, inertia, and proper time is the true foundation of gravitation.}
Geometry provides the symbolic language in which this coherence is expressed—a formal structure encoding the relational order of the physical world. From this standpoint, Einstein’s quest to unify inertia and gravitation finds its natural culmination in the principle of extremal proper time, which encapsulates the mutual determination of matter, motion, and temporal structure within a single dynamical framework.

\begin{acknowledgments}

This work is supported by the Spanish grant PID2021-123903NB-I00
funded by MCIN/AEI/10.13039/501100011033 and by ``ERDF A way of making Europe''. 

\end{acknowledgments}

{

\section*{Appendix: Summary of Previous Works}

The present manuscript builds upon three recent works by Emilio Elizalde and the present author \cite{HE2025,HE2025a,HE2026}. For completeness and to make the discussion more self-contained, we summarize here the main conceptual and technical results that are directly relevant to the developments presented in this paper.

\subsection*{A.1 Dynamical Interpretation of the Equivalence Principle \cite{HE2025}}

In \cite{HE2025}, the historical and conceptual foundations of the Equivalence Principle (EP) were revisited, with particular emphasis on its dynamical content and its role in the early development of General Relativity (GR). The main conclusions can be summarized as follows:

\begin{enumerate}
\item
Free fall in a uniform gravitational field does not necessarily imply the absence of forces. Instead, gravitational and inertial forces can be understood to exactly compensate each other in the sense of D’Alembert’s principle. Within this framework, the inertial force is naturally expressed in terms of proper time as
$$
-m\frac{d^2{\bf r}}{ds^2},
$$
which balances the gravitational interaction along the free-fall trajectory.

\item
The weak equivalence principle (equality of inertial and gravitational masses) admits a consistent dynamical interpretation: a freely falling body does not experience weight because gravitational and inertial contributions precisely compensate. This viewpoint is compatible with Einstein’s original physical intuition and early formulations of the EP.

\item
From this perspective, free-fall motion can be recovered by introducing a suitable modification of the Minkowski metric that incorporates gravitational effects without direct use of Einstein’s field equations. Proper time plays a central role, and gravitation appears as a modulation of its flow.

\item
This procedure naturally leads to an effective metric that reproduces the Newtonian second law within a relativistic framework, thereby providing a dynamical pathway toward the gravitational metric in the weak-field regime.
\end{enumerate}

\subsection*{A.2 Geometric Formulation of Newtonian Cosmology \cite{HE2026}}

In \cite{HE2026}, a gauge-aware geometric formulation of Newtonian cosmology was developed. The main results are the following:

\begin{enumerate}
\item
In light of the Hole Argument \cite{Stachel2014,Davis2019}, General Relativity can be interpreted as a gauge theory under active diffeomorphisms, which emphasizes the relational character of spacetime structure.

\item
Within this framework, the first and second Friedmann equations can be derived using exclusively Newtonian arguments, without assuming the full relativistic formalism.

\item
When formulated geometrically, classical Newtonian cosmology exhibits a relational and gauge-invariant structure analogous to that of GR. In particular, perturbed FLRW dynamics (perturbed first and second Friedmann equations) can be written entirely in terms of geometric objects—such as the Hodge operator, Lie derivatives, and differential forms—without introducing an absolute spacetime background.

\item
The resulting equations coincide with the linearized GR equations in the Newtonian gauge, thus making explicit the connection between Newtonian intuition and relativistic invariant structures.
\end{enumerate}

\subsection*{A.3 Reinterpreting General Relativity and Gravity without Geometry in the Weak Limit \cite{HE2025a}}

In \cite{HE2025a}, the geometric interpretation of General Relativity was critically revisited, emphasizing the following aspects:

\begin{enumerate}
\item
The spacetime metric has an operational meaning: curvature encodes measurable relations among rods, clocks, and light signals, rather than representing a material physical entity.

\item
Although Einstein’s heuristic reasoning (1907–1920) led naturally to the geometric formulation of GR, the weak-field regime can be reconstructed from dynamical considerations without postulating curvature a priori.

\item
Fermat’s principle admits a consistent extension to massive particles: free trajectories extremize proper time, in agreement with Lorentz invariance and relativistic dynamics.

\item
In the weak-field linear regime, this approach reproduces the standard GR solutions in harmonic coordinates, confirming the consistency of a physically motivated derivation of the gravitational metric. Beyond the linear approximation, the natural nonlinear completion emerges through the Ricci tensor.
\end{enumerate}

\subsection*{A.4 Conceptual Takeaways}

Taken together, the three works support the following general picture:

\begin{enumerate}
\item
Gravitational dynamics can be reconstructed from the interplay between inertia and matter without introducing geometric structure as an a priori postulate.

\item
The original formulation of the Equivalence Principle suggests that gravity modifies the flow of time. In particular, the presence of a gravitational field alters proper time, which motivates a corresponding modification of the Minkowski invariant.

\item
By extending Fermat’s principle and using the Galilean law of free fall in a uniform field, one can infer the structure of the invariant interval $ds^2$. This provides guidance for the form of the invariant in a general static field. After implementing Lorentz transformations, the weak-field invariant for moving sources can also be obtained.

\item
The invariant obtained through this methodology coincides with that derived in linearized GR in harmonic coordinates. In this sense, the Ricci tensor naturally appears when enforcing compatibility with energy–momentum conservation and Lorentz invariance, providing a nonlinear completion of the weak-field equations.

\item
Newtonian cosmology admits a consistent geometric formulation, highlighting the continuity between classical, weak-field, and relativistic regimes and clarifying the operational meaning of spacetime quantities.
\end{enumerate}

\vspace{0.3cm}

This summary provides the conceptual and technical background underlying the present manuscript and allows the reader to follow the construction without detailed consultation of the previous works.

}

\end{document}